\documentclass[]{article}

\pdfoutput=1
\usepackage[latin1]{inputenc}
\usepackage{amsmath}
\usepackage{amsfonts}
\usepackage{amssymb}
\usepackage{graphicx}
\usepackage{subcaption}
\usepackage{lpic}
\usepackage{float}
\usepackage{mathtools}
\usepackage{authblk}
\usepackage{multicol}
\usepackage{geometry}

\newcommand{\bk}{\mathbf{k}}
\newcommand{\bx}{\mathbf{x}}
\newcommand{\ud}{\text{d}}
\date{}

\title{Nonlinear wave interaction in coastal \\ and open seas -- deterministic and stochastic theory}

\author[1]{Raphael Stuhlmeier\thanks{raphael.stuhlmeier@plymouth.ac.uk}}
\author[2]{Teodor Vrecica\thanks{teodorv@post.tau.ac.il}}
\author[2]{Yaron Toledo\thanks{toledo@tau.ac.il}}
\affil[1]{Centre for Mathematical Sciences\\
University of Plymouth\\
Drake Circus, Plymouth PL4 8AA\\
United Kingdom}
\affil[2]{School of Mechanical Engineering\\
Tel-Aviv University\\
Tel-Aviv, 69978\\
Israel}

\begin{document}

\maketitle
\begin{abstract}
We review the theory of wave interaction in finite and infinite depth. Both of these strands of water-wave research begin with the deterministic governing equations for water waves, from which simplified equations can be derived to model situations of interest, such as the mild slope and modified mild slope equations, the Zakharov equation, or the nonlinear Schr\"odinger equation. These deterministic equations yield accompanying stochastic equations for averaged quantities of the sea-state, like the spectrum or bispectrum. We discuss several of these in depth, touching on recent results about the stability of open ocean spectra to inhomogeneous disturbances, as well as new stochastic equations for the nearshore. 
\end{abstract}

\section{Introduction}
\subsection{Preliminaries}
The water wave problem as it is understood today is an outgrowth of Newtonian mechanics, and was first cast in the framework of partial differential equations by Leonhard Euler. From its very beginnings, the development of water wave theory went hand in hand with the development of new mathematical tools for treating differential equations. Belying its classical origins, the subject of water waves remains a vibrant area of research to this day.

Much applicable research on ocean waves today is focused on forecasting, which adds a stochastic element to the deterministic equations for the free boundary problem for an inviscid, incompressible fluid. The purpose of the present review is to present some of these ideas, as well as some recent developments, in the subject of deterministic and stochastic wave interaction. Far from being a mathematical abstraction, this body of ideas informs the surfer waiting for a big swell, the marine engineer designing an offshore structure, and the commercial mariners voyaging across the world's oceans and seas.

\subsection{Governing equations}
The governing equations for water waves can by now be found in any textbook on the subject. A clear, modern derivation may be found in Johnson \cite{Johnson1997}. In what follows, the assumptions made of the water will be as follows: it is inviscid (to avoid the Navier--Stokes equations), it is incompressible (so the speed of sound is infinite), and the only restoring force is gravity. The surface tension of water plays an important role for very short waves (periods less than about half a second), but on these scales viscosity also becomes important, and it is expedient to dispense with both. Usually only a single fluid (the water) is considered, and the air above is neglected, allowing a decoupling of the atmosphere from the ocean. This assumption is realistic for the propagation of ocean waves without wind, but must be viewed critically when wind forcing becomes important. One final assumption, less convincing on purely physical grounds, but mathematically important, is that of irrotational flow. While Kelvin's circulation theorem may be invoked to justify this choice, the mathematical convenience of potential flow, i.e.\ replacing the fluid velocity field $\mathbf{u}$ by a potential $\phi$ with $\nabla \phi = \mathbf{u},$ is critical in simplifying all subsequent analysis.

The governing equations with these assumptions are as follows:
\begin{align}
& \Delta \phi = 0 \label{eq:1.2-Laplace}\\
& \eta_t + \phi_x \eta_x + \phi_y \eta_y - \phi_z = 0 \text{ on } z = \eta(x,y,t) \label{eq:1.2-SKBC}\\
& \phi_t + \frac{1}{2}\left(  \phi_x^2 + \phi_y^2 \right) + g \eta = 0 \text{ on } z = \eta(x,y,t)\label{eq:1.2-Bernoulli}\\
& \phi_z  + \phi_x h_x + \phi_y h_y = 0 \text{ on } z = -h(x,y) \label{eq:1.2-BKBC}
\end{align}
Here $g$ is the (constant) acceleration of gravity, $h$ denotes the bottom boundary, and $\eta$ the unknown free-surface. While the bottom may be allowed to vary in space, it will be fixed in time -- so we cannot consider, for example, the generation of a tsunami by an earthquake.

Since all nonlinearity is contained in the kinematic surface \eqref{eq:1.2-SKBC} and bottom \eqref{eq:1.2-BKBC} boundary conditions, and the Bernoulli condition \eqref{eq:1.2-Bernoulli}, upon linearization this problem becomes a standard exercise in solving Laplace's equation.

Allowing for bathymetry, the linear problem takes the form 
\begin{align}
& \Delta \phi = 0, \label{eq:1.2-Laplace-linear}\\
& \phi_{tt} + g \phi_z = 0 \text{ on } z = 0, \label{eq:1.2-CSBC}\\
& \phi_z  + \phi_x h_x + \phi_y h_y = 0 \text{ on } z = -h(x,y), \label{eq:1.2-BKBC-lin}
\end{align}
where the surface boundary conditions have been combined to eliminate $\eta$ from the problem. Equation \eqref{eq:1.2-BKBC-lin} simplifies to $\phi_z = 0$ on $z=-h$ for constant depth, resulting in the Laplace equation on a horizontal strip, or, for infinite depth, a half space.

 It suffices here to record a few main results: travelling wave solutions in constant depth have the form $\exp(i(\bk \mathbf{x} - \omega(\bk)t)),$ where the relationship between $\bk$ and $\omega(\bk)$ is given by 
\begin{equation} \label{eq:1 linear dispersion relation}
\omega(\bk)^2 =  g |\bk| \tanh( |\bk| h).
\end{equation}
When the depth varies it makes sense to define a local wavenumber and frequency -- in general we may have $\bk=\bk(\bx,t), \, \omega = \omega(\bx,t).$ Thus, we have travelling wave solutions $\exp(i S(\bx,t))$ for a phase-function $S,$ and through this define $\bk = \nabla S$ and $\omega = \partial S/\partial t.$

\section{Nonlinear waves \& interaction}
\label{sec:Nonlinear waves and interaction}

The theory of nonlinear water waves was historically first treated by perturbation expansions, dating back to the work of Stokes in the mid $19^{\text{th}}$ century. The procedure starts by expanding $\phi$ and $\eta$ in \eqref{eq:1.2-Laplace}--\eqref{eq:1.2-BKBC} in terms of a small factor $\varepsilon,$ and transferring the free surface from $z = \varepsilon \eta$ to $z=0$ by a Taylor expansion. One may then solve \eqref{eq:1.2-Laplace-linear}--\eqref{eq:1.2-BKBC-lin} first for terms of order $O(1),$ the solution of which then appears as an inhomogeneity in the equations for $O(\varepsilon^1),$ and so on. The algebra quickly becomes cumbersome, particularly for finite depth, and if more than one wave-train is involved.

It is easier to start with simpler equations, and a good introduction is furnished by Whitham \cite[Sec.\ 15.6]{Whitham1974}. Assume for the moment that we have a nonlinear, dispersive equation of the form 
\[ \phi_{tt} + \mathcal{L}_x(\phi) = \varepsilon \mathcal{N}(\phi), \]
where $\varepsilon$ is some small parameter, $\mathcal{L}_x$ is a linear differential operator involving spatial ($x$) derivatives, and $\mathcal{N}$ is some nonlinear operator. The linearised problem, for $\varepsilon = 0,$ has travelling wave solutions of the form $\exp(i(k {x} - \omega(k)t)),$ as above, where $\omega$ depends on the operator $\mathcal{L}_x.$ 

In the linear problem, the sum of two plane-wave solutions $\exp(i(k_1x - \omega_1 t))$ and $\exp(i(k_2x - \omega_2 t))$ is again a solution. However, if $\mathcal{N}$ contains a term $\phi^2$, then a sum of two solutions results in the product $\exp(i((k_1+k_2)x - (\omega_1 + \omega_2)t))$ on the right-hand side. If the nonlinearity is cubic, then three travelling waves can combine on the right-hand side to $\exp(i((k_1+k_2+k_3)x - (\omega_1 + \omega_2 + \omega_3)t)).$ These terms act as a forcing for the linear equation $\phi_{tt} + \mathcal{L}_x(\phi).$ Just as in the classical theory of forced linear oscillators, the critical phenomenon is \emph{resonance,} when the frequency of the forcing matches that of the unforced system. 

Accounting for all possible sums and differences, we see that resonances for quadratic nonlinearities involve three waves (one from the left-hand side of the equation, and two from the right)
\begin{equation} \label{eq:2 triad resonance}
\begin{cases}
\pm \bk_1 \pm \bk_2 \pm \bk_3 = 0,\\
\pm \omega(\bk_1) \pm \omega(\bk_2) \pm \omega(\bk_3) = 0,
\end{cases}
\end{equation}
and those for cubic nonlinearities involve four waves
\begin{equation} \label{eq:2 quartet resonance}
\begin{cases}
&\pm \bk_1 \pm \bk_2 \pm \bk_3 \pm \bk_4 = 0,\\
&\pm \omega(\bk_1) \pm \omega(\bk_2) \pm \omega(\bk_3) \pm \omega(\bk_4) = 0.
\end{cases}
\end{equation}

These very relations also arise in the expansion of the water wave problem in a small parameter (like the wave slope $ka$),  where the dispersion relation is given by \eqref{eq:1 linear dispersion relation}. In the limit of infinite depth ($h\rightarrow\infty$) \eqref{eq:1 linear dispersion relation} reduces to $\omega(\bk) = \sqrt{g |\bk|},$ and \eqref{eq:2 triad resonance} cannot be fulfilled nontrivially. For this limit, \eqref{eq:2 quartet resonance} can be fulfilled only for two $+$ and two $-$ signs in both equations. 

\begin{figure}
\begin{centering}
\includegraphics[scale=0.25]{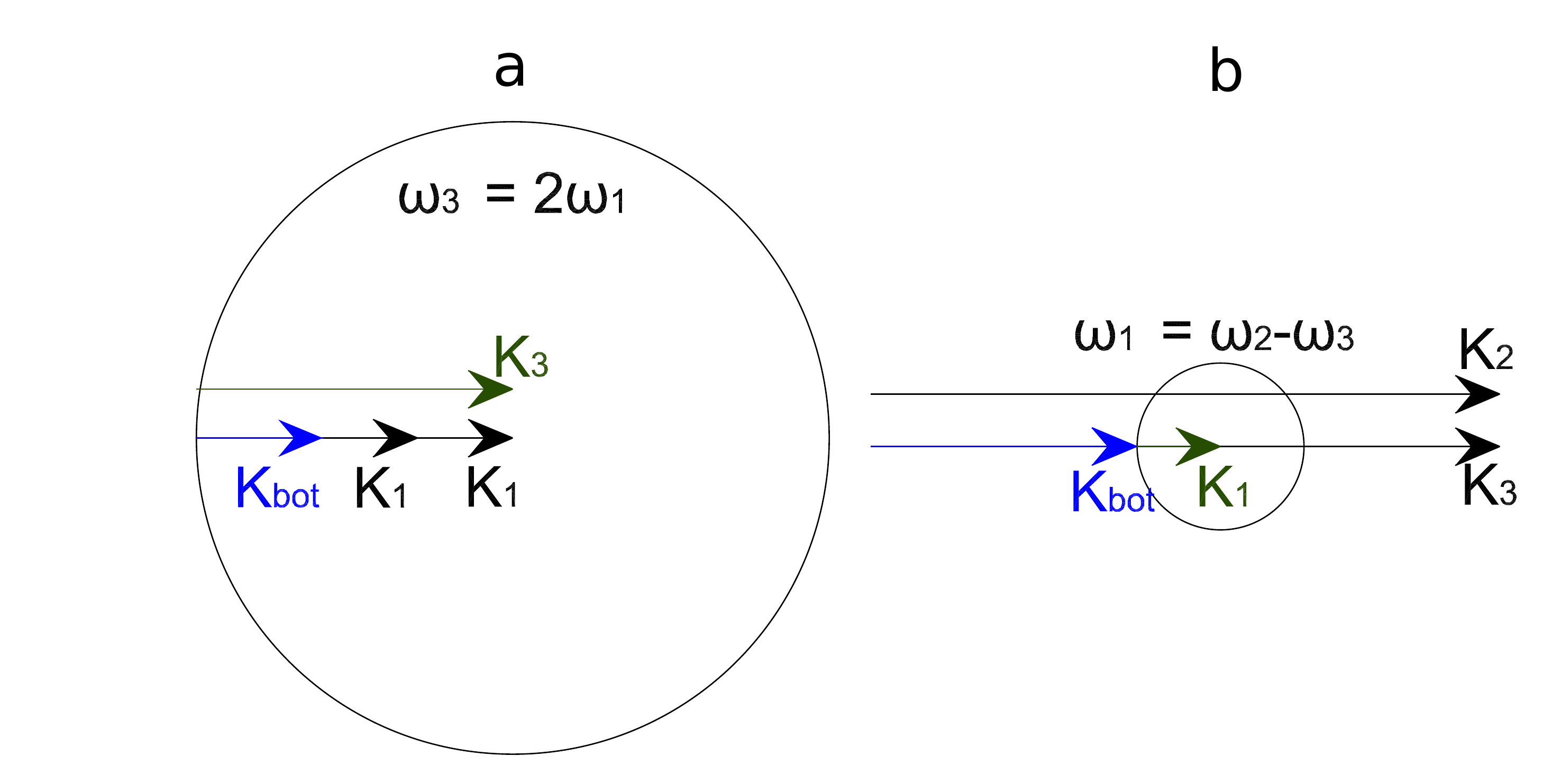}
\par\end{centering}
\begin{centering}
\includegraphics[scale=0.25]{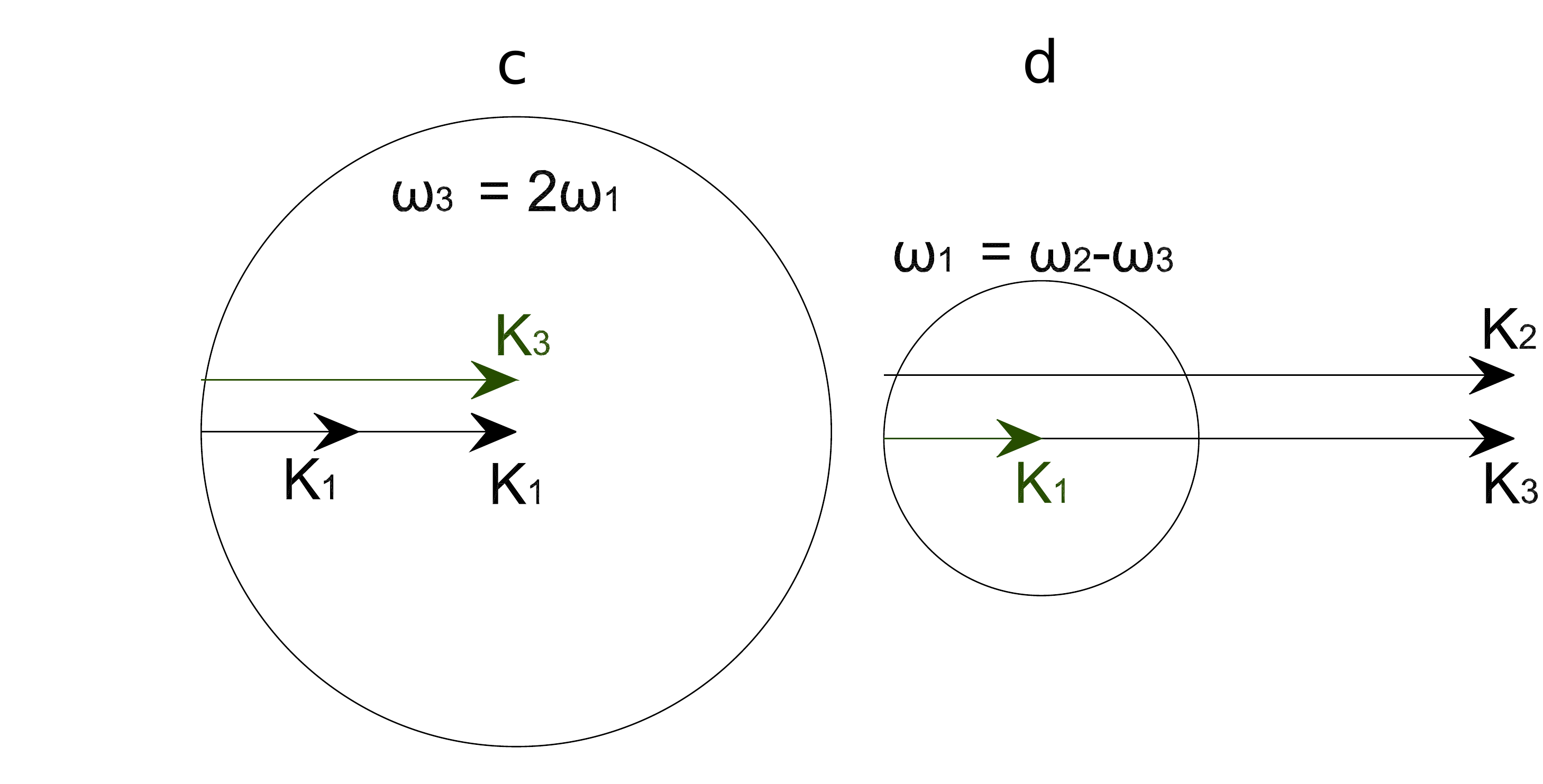}
\par\end{centering}
\caption{\label{fig:Resonance}Resonance conditions in intermediate (a and
b) and shallow water (c and d). Superharmonic (a and c), and subharmonic
interactions are shown (b and d). The $\bk_{bot}$ represents the bottom
component which closes the Bragg III resonance condition.}
\end{figure}

The opposite extreme, of shallow water ($ |\bk| h \ll 1$) means that $\omega = |\bk| \sqrt{g h},$ whereby already \eqref{eq:2 triad resonance} can be fulfilled, provided the signs are not all the same and all wave components propagate in the same direction (see figures \ref{fig:Resonance}c and d). Due to the lack of dispersivity (i.e.\ all wave frequencies propagate with the same celerity) \eqref{eq:2 quartet resonance} is also fulfilled, as are the resonance conditions for any higher order interaction (see \cite{VT2016}).  Nevertheless, the evolution equations are almost always limited to $O\left(ka\right)^{2}$ for finite water depth. While higher order nonlinear terms may be relevant for high wave steepness, their treatment is extremely cumbersome and will not be considered further.  

As waves propagate from deep to shallow water, they are transformed
due to bottom changes. In intermediate
waters, the changing depth induces a change of the wavenumbers through
the dispersion relation \eqref{eq:1 linear dispersion relation}. This alone does not enable the closure of a triad resonance condition and an additional component is required. This component may be supplied by a bottom-induced free-surface interference, which does not satisfy the dispersion relation.

 Assume the bottom profile $h$ is decomposed into a sum of sinusoids, then any bottom wavelength can act as a fourth wavenumber component ($\bk_{bot}$) with a still ($\omega(\bk_{bot})=0$) disturbance on the surface. This allows the closure of  \eqref{eq:2 quartet resonance} in what is known as a class III Bragg-type resonance condition
\begin{equation} \label{eq:triad resonance}
 \begin{cases}
  \pm \bk_{1}\pm \bk_{2}\pm \bk_{3}=\pm \bk_{bot}\\
 \pm \omega(\bk_1) \pm \omega(\bk_2) \pm \omega(\bk_3) = 0,
 \end{cases}
\end{equation}
 with a triad of waves. { This closure can be represented graphically. Figure \ref{fig:Resonance}(a--b) and (c--d) shows this closure for superharmonic and subharmonic 1D interactions in intermediate water and shallow water conditions, respectively. In the 2D case,  bottom components can close resonance with any direction of the third wave $\bk_3$. The circles represent the wavenumber $\bk_3$ satisfying the closure in all possible directions. Depending on its direction, a bottom component $\bk_{bot}$ that satisfies the class III closure should connect its origin on the circle with the origin of the other two waves (see \cite[Sec.\ 3.3]{Liu&yue1998jfm}). }

\section{Deterministic evolution equations}

\subsection{The Zakharov equation for constant depth}
\label{ssec:Zakharov equation}
It is expedient to take a consistent Fourier analysis perspective with equations \eqref{eq:1.2-Laplace}--\eqref{eq:1.2-BKBC}, under the assumption of constant depth $h,$ rather than considering interaction of individual wave-trains via perturbation theory, as was done in \cite{Tick1959, Phillips1960, Longuet-Higgins1962c}. This approach leads after considerable labour to the Zakharov equation, first derived in \cite{Zakharov1968}. 

In terms of a complex amplitude $B(\bk,t),$ related at lowest order to the free-surface elevation $\eta$ by 
\begin{equation} \label{eq:eta definition zaharov}
\hat{\eta}(\bk,t) = \sqrt{\frac{\omega(\bk)}{2g}} \left( B(\bk,t)e^{i\omega(\bk)t} + \text{c.c.} \right) 
\end{equation}
(where ``c.c." denotes the complex conjugate, and $\, \hat{}$ denotes the ($\bx\rightarrow\bk$) Fourier transform)
 the Zakharov equation has the following form 
\begin{align}
\label{eq:3-Zakharov equation}
i \frac{\partial B(\bk,t)}{\partial t} = \iiint_{-\infty}^\infty T_{0123} B_1^*(t) B_2(t) B_3(t)\delta_{0+1-2-3}
 e^{i(\Delta_{0+1-2-3})t} \ud \bk_1 \ud \bk_2 \ud \bk_3 
\end{align}
where $\delta_{0+1-2-3} = \delta(\bk+\bk_1-\bk_2-\bk_3)$ is the delta-distribution, $\Delta_{0+1-2-3} = \omega(\bk)+\omega(\bk_1)-\omega(\bk_2)-\omega(\bk_3),$ and $T_{0123}=T(\bk,\bk_1,\bk_2,\bk_3)$ is a very lengthy interaction kernel (see \cite{Krasitskii1994}). For brevity we have denoted by $B_j(t)$ the complex amplitude $B(\bk_j,t),$ and the superscript $*$ denotes a complex conjugate. While a detailed discussion of the Zakharov equation can be found elsewhere (see, e.g.\ \cite[Sec.\ 14]{Mei2005}) it is important to appreciate that equation \eqref{eq:3-Zakharov equation} arises from a multiple-scale \emph{ansatz} for the full third-order Fourier-space problem, and so captures terms with cubic nonlinearities. In terms of the small wave steepness $\varepsilon,$ the time $t$ in this equation is related to physical time $T$ by $t = \varepsilon^2 T,$ which is the same as the slow time-scale for the evolution of the envelope in the nonlinear Schr\"{o}dinger equation \cite[Eq.\ (4.2)]{Johnson1997}. This long time scale must be borne in mind for all subsequent results.

For computational implementation (and even analytic studies of systems with few waves) it is inevitable that \eqref{eq:3-Zakharov equation} must be discretized. Making an ansatz that $B(\bk,t) = \sum_{i = 1}^{N} B_i(t) \delta(\bk - \bk_i),$ i.e.\ that the complex amplitudes can be written as a sum of generalized functions, and integrating over a ball centered around $\bk_n$ yields
\begin{equation}
\label{eq:3-DZE}
i \frac{\ud B_n(t)}{\ud t} = \sum_{p,q,r=1}^N T_{npqr} \delta_{n+p-q-r} e^{i \Delta_{n+p-q-r}t} B_p^*(t) B_q(t) B_r(t)
\end{equation}
where 
\begin{equation}
\delta_{n+p-q-r} = \begin{cases}
1 \text{ for } \bk_n + \bk_p = \bk_q + \bk_r \\
0 \text{ otherwise}
\end{cases}
\end{equation}
now denotes a Kronecker delta-function. Note that other approaches to discretization have been taken, for example by Rasmussen \& Stiassnie \cite{Rasmussen1999} or Gramstad \emph{et al} \cite{Gramstad2011}, although their applications largely await further study.

The simplest solution to \eqref{eq:3-DZE} is obtained when only a single wave is present, where with the identity (valid for deep-water)
\begin{equation} \label{eq:3-T0000}
T(\bk,\bk,\bk,\bk) = \frac{|\bk|^3}{4 \pi^2}
\end{equation}
the third-order Stokes' correction is recovered as expected (see \cite[Sec.\ 14.5]{Mei2005}). Similarly, if only two collinear waves are present, and we denote these by scalar wavenumbers $k_1, \, k_2$ the identity
\begin{equation}
T(k_1,k_2,k_1,k_2) = \begin{cases}
{k_1 k_2^2}/{(4 \pi^2)} \text{ if } k_2 < k_1 \\
{k_1^2 k_2}/{(4 \pi^2)} \text{ if } k_2 \geq k_1 
\end{cases}
\end{equation}
(see \cite[Eq.\ (4.18)]{Zakharov1992}) can be used to establish the mutual Stokes' correction of two wave-trains in deep water, as in \cite[eq.\ (2.11)]{Longuet-Higgins1962c}. In finite depth the kernels are more involved, and do not yield such compact expressions. In particular, work of Janssen \& Onorato \cite{JanssenOnorato2007} first pointed out the problem of non-unique limits for the finite-depth kernel $T(\bk,\bk,\bk,\bk),$ which was later studied in depth, including for kernels of the form $T(\bk,\bk_1,\bk,\bk_1),$ by Stiassnie \& Gramstad \cite{StiassnieGramstad2009} and Gramstad \cite{Gramstad2014}. A significant consequence of \cite{JanssenOnorato2007} is that modulational instability was shown to disappear for $k_0 h<1.363,$ where $k_0$ is the carrier wavenumber, and $h$ the (constant) water depth.

In fact, in Zakharov's \cite[Eq.\ (2.3)]{Zakharov1968} derivation of the eponymous equation, it was not the endpoint of his analysis, but rather a step towards the derivation of the nonlinear Schr\"{o}dinger equation (NLS), which itself was used to study the stability of deep-water waves to perturbations. Having moved from a description in physical variables $(x,y,z,t)$ of fluid motion via the PDEs \eqref{eq:1.2-Laplace}--\eqref{eq:1.2-BKBC}, to a third-order simplification (written in terms of variables defined only at the free surface, thus eliminating $z$) in $(k_x,k_y,t)$ via the integro-differential Zakharov equation \eqref{eq:3-Zakharov equation}, it is possible to make further restrictions so that an inverse Fourier-transform can be carried out.

The central assumption needed to derive NLS is that all interacting waves are clustered about a single wavenumber, say $\bk_0,$ an assumption usually referred to as ``narrow-bandwidth". This is less apparent when deriving the NLS from the governing equations by perturbation theory \cite[Sec.\ 4.1.1]{Johnson1997}, but implicit also in this formulation. On this basis, the Zakharov kernel in \eqref{eq:3-Zakharov equation} is replaced by the kernel $T(\bk_0,\bk_0,\bk_0,\bk_0)$ of \eqref{eq:3-T0000}, and the frequency $\omega(\bk)$ is expanded in a Taylor series about $\omega(\bk_0).$ These two steps allow the inverse Fourier transform to be carried out, and lead to the NLS in much the same way that Zakharov \cite[Eq.\ (2.7)ff]{Zakharov1968} first outlined. 

\subsection{Shallow water and varying depth}
\label{ssec:Shallow water}
We have noted in Section \ref{sec:Nonlinear waves and interaction} that no triad resonance is possible in finite, constant depth. The nature of the perturbation arguments involved implies that quadratic terms are associated with faster time-scales (and larger corrections) than cubic terms, which are in turn more significant than quartic terms -- assuming all the necessary interactions are allowed by the dispersion relation. Thus the Zakharov equation \eqref{eq:3-Zakharov equation} contains only cubic terms, the non-resonant quadratic terms having been eliminated, and the resonant quartic terms being neglected (indeed, it would be more accurate to refer to \eqref{eq:3-Zakharov equation} as the reduced Zakharov equation, see \cite{Krasitskii1994} or \cite{Stiassnie1984b} for the related fourth-order equations in constant depth). For waves in the deep water of the open ocean, this is perfectly satisfactory, but once waves enter coastal environments new equations are needed to capture the effects of a changing bathymetry.

In the shallow water limit, waves become non-dispersive and are able to close exact triad resonances. In real seas waves will almost always tend to steepen and break before reaching the shallow water limit allowing for exact resonance. Nevertheless,  breaking does not extract all of the wave energy immediately. It is a gradual process, in which breaking and nonlinearity are coupled. Due to the inherent complexities of wave breaking (only empirical terms for breaking exist), we simply note its importance and include a general dissipation term in the equations. Hence, no nonlinear shoaling examples that include breaking are presented in this chapter. However, even without reaching exact resonance, the nonlinear triad interactions are still of great importance in coastal areas. It is shown here that wave propagation even over mildly varying bathymetry \cite{Liu&yue1998jfm} leads to quasi-resonance and significant transformations of wave spectra. The subsequent deterministic model equations are often called mild slope-type equations \cite{AgnonEtAl1993}.

The derivation of the linear mild slope equation is based on the equations \eqref{eq:1.2-Laplace-linear}--\eqref{eq:1.2-BKBC-lin}. If the bed is flat, i.e.\ $h$ is constant, the Laplace equation can be separated, and the vertical component of the velocity potential is 
\begin{equation}\label{vertical structure}
f(z) = \frac{\cosh k(z +h)}{\cosh kh}.
\end{equation}
The key to the mild slope equation is assuming this functional form for the $z$-dependent part of the solution, even when the bottom is not of constant depth. Thus the solution to \eqref{eq:1.2-Laplace-linear}--\eqref{eq:1.2-BKBC-lin} is written ${\phi}(x,y,z) = -ig\eta(x,y) \omega^{-1} f(z),$ (here $\phi$ denotes the time-harmonic velocity potential) and the explicit depth-dependence is integrated out using Green's identity:
\[ \int_{-h}^0 \left( f_{zz} {\phi} - f {\phi}_{zz}\right) \ud z = \left[ f_z {\phi} - f{\phi}_z \right]_{-h}^0. \]
For a flat bed $-h=\text{const.}$ the right-hand side vanishes, and we are left with a Helmholtz equation. Otherwise, a varying bed $h = h(x,y)$ gives rise to the mild slope equation when terms $O(\nabla_h h)^2$ and $O(\nabla_h^2 h)$ are neglected:
\begin{equation}\label{eq:MSE}
\nabla_h \cdot ( a \nabla_h \hat{\phi} ) + k^2 a \hat{\phi} = 0,
\end{equation}
where $a(x,y) =g \int_{-h(x,y)}^0 f(z)^2 dz.$ Here, $\nabla_{h}=(\partial_x,\partial_y)$ is the horizontal gradient and $\hat{\phi}$ is a single harmonic of the velocity potential on the linearised free surface $(z=0)$. In fact, $a$ is exactly the product of the phase velocity and the group velocity, $a = \omega/k \cdot d\omega/dk = C_p \cdot C_g.$ More details can be found in \cite[Ch.\ 3.5]{Mei2005}. Note that upon retaining higher order bottom terms one can derive the more accurate Modified MSE (see \cite{ChamberlainPorter1995}).

Nonlinear mild-slope evolution equations can also be derived from the governing equations \eqref{eq:1.2-Laplace}--\eqref{eq:1.2-BKBC}, with the vertical structure of velocity potential either assumed to be that of a free wave as in equation \eqref{vertical structure}  \cite{AgnonEtAl1993,KaihatuKirby1995}, or expanded as a Frobenius series \cite{BredmoseEtAl2005ejmbf,VT2016}, with the latter giving better accuracy in the nonlinear part. The general form, written in terms of the surface velocity potential for a given harmonic $p$ is defined as
\begin{equation}
\nabla_{h}^2\hat{\phi}_{p}+\frac{\nabla_{h}\left(C_{p}C_{g,p}\right)\cdot\nabla_{h}\hat{\phi}_{p}}{C_{p}C_{g,p}}
+k_{p}^{2}\hat{\phi}_{p}=\textrm{NL}_{p},\label{eq:NLMSE}
\end{equation}
where $C_{p}$ and $C_{g,p}$ are wave celerity and group velocity for harmonic $p$, while $\textrm{NL}_{p}$ is the nonlinear triad term, which closes an exact resonance in frequency for harmonic $p$.

In order to evaluate the evolution of the wave field, the model is often parabolized or hyperbolized (see \cite{Radder1979}), by assuming a progressive wave of the form
\begin{equation}
\eta_{p,l}=a_{p,l}e^{-i\left(-k_{l}^{y} y-\int_{0}^{x}k_{p,l}^{x'}dx'+\omega_{p}t\right)},\label{eq:Par}
\end{equation}
similar to \eqref{eq:eta definition zaharov} with $k_{p,l}^{x}$ and $k_{l}^{y}$ representing the $x$- and $y$-components of the wave number vector respectively. The $l$-index relates to the discretisation in the lateral direction, where no bottom changes are assumed. This allows the direct satisfaction of the resonance closure in the lateral direction and a decoupling between directional components of each harmonic.

The relation between $\eta_{p}$ and  $\hat{\phi}_p$ can be found using a Taylor series expansion of \eqref{eq:1.2-Bernoulli} about  $z=0$. Note that one should retain $O(\varepsilon^2)$ terms in this relation in order to remain consistent with the equation's order (see \cite{EldeberkyMadsen}).   Based on \cite{BredmoseEtAl2005ejmbf} and \cite{VT2016} the deterministic wave evolution equation for the Fourier amplitude  of the surface elevation $a_{p,l},$ with constant lateral wavenumber $k_{l}^{y},$ is defined as
\begin{align}
& \frac{1}{C_{g,p}}\frac{\partial a_{p,l}}{\partial t}+\frac{\partial a_{p,l}}{\partial x}+\frac{1}{2C_{g,p}}\frac{\partial C_{g,p}}{\partial x}a_{p,l}+D_{p,l}a_{p,l}\nonumber \\
& = -\varepsilon i
\sum_{\mathclap{\substack{{u=\max\{l-M,-M\}}\\{s=p-N}}}}^{\mathclap{\substack{{s=N}\\{u=\min\{l+M,M\}}}}}
W_{s,p-s,u,l-u}a_{s,u}a_{p-s,l-u}e^{-i\int\left(k_{s,u}^{x}+k_{p-s,l-u}^{x}-k_{p,l}^{x}\right)dx}.\label{eq:EvolutionElevation}
\end{align}
Here, the $W$-term is the nonlinear interaction kernel defined in Bredmose et al \cite{BredmoseEtAl2005ejmbf} and Vrecica \& Toledo \cite{VT2016} for cases without and with dissipation, respectively. $D_{p,l}$ can describe
a linear damping or forcing term, while $t = \varepsilon^2 T$ represents a slow time evolution,
which is typically on a different scale than the spatial evolution $x = \varepsilon X,$ for $T$ and $X$ physical time and space variables (cf.\ the comments after \eqref{eq:3-Zakharov equation} in Sec.\ \ref{ssec:Zakharov equation}). It appears when one allows the potential in the mild-slope type equation to vary slowly in time on top of its harmonic behaviour.

Typically wave reflection and nonlinear generation in the backwards
direction are second order effects, and are not considered further. They
can become significant under certain conditions, and to account for
them it is possible either to solve the nonlinear elliptic MSE given in \eqref{eq:NLMSE} or to create two coupled evolution equations -- one for forward propagating waves and the other for backward propagating waves in the same manner as in the linear case (see \cite{Radder1979}).

\subsection{Explanation of nonlinear energy transfer using spring-mass allegory}

One way to think about the wave resonance phenomenon is via an analogy to
oscillating mass-spring systems. The linear part of \eqref{eq:NLMSE}, upon redefinition of  $\hat{\phi}_p$ and $k_p$, takes the form of a Helmholtz equation (see \cite{Radder1979}), which in one dimension becomes a simple harmonic oscillator. 
Imagine a set of $N$ oscillating spring-mass systems, related to $N$ spectral frequency bins. Softer springs (small spring constant $k_p$) are in lower harmonics, and as the frequency increases the springs become stiffer (larger $k_p$). When the problem is linear, these systems are decoupled, but nonlinear terms couple each spring system (spectral bin) to the oscillation of other spring systems.

 The nonlinear part of \eqref{eq:NLMSE}, which relates to combinations of waves that already satisfy the resonance condition in $\omega$,  acts as a forcing term on the mass-spring system, as explained schematically in Section \ref{sec:Nonlinear waves and interaction}. These forcing terms are generally small in magnitude, compared to the total energy of the system. Non-resonant forcings (i.e., ones that do not close the resonance condition in $k$) will cause the system to oscillate slightly at the frequency of forcing (bound wave). If the forcing matches the spring's natural frequency (i.e., the resonance condition in wavenumber is met), a resonance is reached and a significant amount of energy transfers to the related spectral bin.

\section{Stochastic evolution equations}
\subsection{Introduction}
The deterministic equations given above seem to provide fertile material for modelling the sea. Under the assumption that the waves are not too steep (in particular, not breaking), so as to remain in a weakly nonlinear regime, and that there are no further forces, it seems that if suitable initial conditions can be supplied the subsequent evolution could be solved for numerically. If we are able to measure a sea-surface, and to conclude that it is composed of Fourier modes $\bk_i$ with given amplitudes $a_i,$ suitable initial conditions for the discrete Zakharov equation \eqref{eq:3-DZE} consist of specifying $B_i(t=0).$ 

If one is interested in average quantities of the sea-state, like the energy, it becomes necessary to develop new evolution equations, in particular since we cannot accurately specify initial conditions for all situations of interest. In particular, the wave phases are found to be essentially uniformly distributed between $(0,2\pi].$ Underlying this approach is the idea that the free surface $\eta(x,y,t)$ (or the complex amplitudes in our deterministic equations) is a stochastic process. The perspective taken here is that the temporal evolution of any realization is governed by a given deterministic equation -- for example the Zakharov equation \eqref{eq:3-Zakharov equation}.  This is a suitable viewpoint for waves at sea, but in a wave-tank it may be more appropriate to consider a spatial evolution equation instead (see, for example, Shemer \& Chernyshova \cite{Shemer2017}). Our assumption also means that no random forcing by the wind, or the like, plays a role in the evolution of the wave field. 

The energy density spectrum, based on linear theory, rests on an underlying assumption of homogeneity of the sea state. This is a prerequisite for sensible measurements (see \cite[Sec.\ 3.5, App.\ A \& App.\ C]{Holthuijsen2008}, or \cite[Sec.\ 9]{Kinsman1984}) and is a convenient starting point for assumptions that are made in the equations for the temporal evolution of energy spectra. In practical measurements of waves, stationarity (or homogeneity) means that the conditions are unchanged for the duration of the measurement (or over the space being measured). For example, it clearly makes no sense to average two measurements of the sea-surface elevation if one is windward and the other leeward of an island.

While the literature on nonlinear stochastic evolution equations is vast, it is worth pointing the reader to some of the resources with a bearing on water wave theory. The stochastic approach to ocean waves was initiated by Pierson (see \cite{Pierson1955}) in the 1950s, and an account of the field up to the mid 1960s is found in the engaging work of Kinsman \cite{Kinsman1984}. A general perspective on weakly nonlinear dynamics, also touching on other fields, is provided by Zakharov et al \cite{Zakharov1992a} and Nazarenko \cite{Nazarenko2011}, while Janssen \cite{Janssen2004} places this theory firmly in the context of modern wave forecasting. A particularly clear account of many aspects of nonlinear and random waves may also be found in a book chapter by Trulsen \cite{Trulsen9999}.

\subsection{Stochastic evolution equations for deep water}
\label{subsec:stochastic evolution equations}
Historically, the first treatment of the evolution of a spectrum of surface waves dates back to Hasselmann \cite{Hasselmann1962}, shortly after the discovery of resonant interaction theory for surface waves in deep water by Phillips \cite{Phillips1960}. It is simpler to start with the later work of Longuet-Higgins \cite{Societyc}, whose point of departure is the 2D NLS in deep water -- as mentioned above, this can be derived from the Zakharov equation.
\subsubsection{Narrow-band equations}
\label{ssec:Narrow-band equations}
We start, following Longuet-Higgins, with the scal\-ed form of the 2D NLS
\begin{equation} \label{eq:2D NLS} 2i A_\tau = \frac{1}{4}(A_{xx} - 2 A_{yy}) + |A|^2A, \end{equation}
with $A=A(x,y,\tau)$ the envelope amplitude, $x \text{ and } y$ slow spatial variables, and $\tau = \varepsilon^2 T$ a slow time. Two approaches are possible, in either physical or in Fourier space, and we explore the former first -- the main ideas are identical for both, and can be found, for example, in \cite[Sec.\ 2]{Zakharov1992a}

Step 1: write \eqref{eq:2D NLS} at a point $\mathbf{x}_1 = (x_1,y_1),$ and multiply it by $A^*(\mathbf{x}_2)=A^*(x_2,y_2),$ where $*$ stands for a complex conjugate. Step 2: Subtract the complex conjugate of the resulting equation from itself. Assume that the envelope amplitudes $A(\bx,\tau)$ are stochastic processes, such that each realization is governed by the deterministic NLS \eqref{eq:2D NLS}. Step 3: take averages (expected values) of the equation from step 2 to obtain
\begin{align} \nonumber 
2i \frac{\partial}{\partial \tau} \langle A(\bx_1) A^*(\bx_2) \rangle &= \frac{1}{4} \left( \frac{\partial^2}{\partial x_1^2} - \frac{\partial^2}{\partial x_2^2} \right) \langle A(\bx_1)A^*(\bx_2) \rangle  \\ 
& - \frac{1}{2} \left( \frac{\partial^2}{\partial y_1^2} - \frac{\partial^2}{\partial y_2^2} \right) \langle A(\bx_1) A^*(\bx_2) \rangle \label{eq:proto-Alber-eq} \\ 
&+ \langle A(\bx_1)A^*(\bx_1)A(\bx_1)A^*(\bx_2) \rangle - \langle A(\bx_2)A^*(\bx_2)A(\bx_2)A^*(\bx_1) \rangle. \nonumber
\end{align}
At this point, further progress depends on stochastic assumptions made for $A.$ The principal obstacle is to treat the fourth-order averages appearing on the right-hand-side of \eqref{eq:proto-Alber-eq}. Assuming that the process $A$ is close to Gaussian, and has zero mean, allows the decomposition
\[ \langle A(\bx_1)A^*(\bx_1)A(\bx_1)A^*(\bx_2) \rangle = 2 \langle A(\bx_1)A^*(\bx_1) \rangle \langle A(\bx_1)A^*(\bx_2) \rangle,  \]
where higher order cumulants have been discarded entirely
We can factorize the differential operators on the right-hand side of \eqref{eq:proto-Alber-eq} by introducing $\mathbf{R} = (r_x,r_y) = \bx_1 - \bx_2$ and $\mathbf{X} = (X,Y) = \frac{1}{2}(\bx_1 + \bx_2),$ and with $C(\mathbf{R},\mathbf{X}) := \langle A(\bx_1) A^*(\bx_2) \rangle$ rewriting \eqref{eq:proto-Alber-eq} as
\begin{equation} \label{eq: deep-water Alber}
2i \frac{\partial}{\partial \tau} C = \frac{1}{2} \frac{\partial^2}{\partial r_x \partial X} C - \frac{\partial^2}{\partial r_y \partial Y} C + 2 C \langle A(\bx_1) A^*(\bx_1) \rangle - 2C \langle A(\bx_2) A^*(\bx_2) \rangle.
\end{equation}
This is, up to scaling, the deep-water analogue of Alber's equation \cite[Eq.\ (3.7)]{Societyb}. 

If, in addition, we assume that $A$ is homogeneous in (physical) space, then averages must be invariant under translation, i.e.\ the autocorrelation $C$ must depend only on spatial separation $\mathbf{R},$ and not on the average position $\mathbf{X}.$ Employing this homogeneity condition in \eqref{eq: deep-water Alber} gives $\partial C/\partial \tau = 0$ at this order, as all terms on the right-hand side of \eqref{eq: deep-water Alber} vanish. To proceed with a statistically homogeneous theory, the lowest order decomposition of the fourth-order terms $\langle A(\bx_1)A^*(\bx_1)A(\bx_1)A^*(\bx_2) \rangle$ must be corrected, by using the product rule and \eqref{eq:2D NLS} in considering $\partial/\partial \tau \langle A(\bx_1)A^*(\bx_1)A(\bx_1)A^*(\bx_2) \rangle.$ In addition, higher-order cumulants and moments will have to be retained and treated accordingly (see \cite{KimPowers1979}).

Longuet-Higgins \cite{Societyc} pursued exactly such an aim, albeit in Fourier space, substituting 
\[ A = \sum_n a_n(\tau) e^{i (\lambda_n x + \mu_n y - \omega_n \tau)}\]
into \eqref{eq:2D NLS}, and using the dispersion relation $\omega_n = -\frac{1}{8} (\lambda_n^2 - 2 \mu_n^2).$ Equating coefficients, and denoting $\bk_i = (\lambda_i,\mu_i),$ he found \cite[Eq.\ 4.3]{Societyc}
\begin{equation}
\label{eq:k-space NLS}
2i \frac{d a_n}{d \tau} = \sum_{p,q,r} a_p a_q a_r^* e^{i(\omega_p + \omega_q - \omega_r - \omega_n)\tau} \delta(\bk_p + \bk_q - \bk_r - \bk_n), 
\end{equation}
which is formally identical (except for a factor of 2) with \eqref{eq:3-DZE} when the kernel is  taken as a constant\footnote{However, the $\omega_i$ satisfy the dispersion relation of the NLS rather than the linear deep-water dispersion relation in \eqref{eq:3-DZE}.}. To now derive a stochastic evolution equation, follow the three steps above in $\bk-$space: multiply \eqref{eq:k-space NLS} by $a_m^*,$ subtract the complex conjugate equation, and take averages. Homogeneity in physical space now means a lack of correlation of Fourier modes \cite[Eq.\ (11.75)]{Papoulis2002}, so that $\langle a_n a_m^* \rangle = C_n \delta(n - m),$ while the quasi-Gaussian closure remains unchanged. Finally, this leads to \cite[Eq.\ 4.10]{Societyc}, a discrete, narrow-band kinetic equation.

Thus the same deterministic equation has yielded two different stochastic evolution equations, depending on whether or not statistical homogeneity is imposed. Homogeneity means that the energy density (as measured by our correlators) does not change to lowest order, so that the time-scale of evolution is longer. Note in \eqref{eq: deep-water Alber} that the rate of change of the energy density $C$ is proportional to $C,$ whereas in the homogeneous case \cite[Eq.\ (4.10)]{Societyc} we find $dC/dt \propto C^3.$ Thus a homogeneous sea-state can be expected to change only slowly due to nonlinear interactions, except possibly when perturbed by inhomogeneous disturbances.

\subsubsection{Stability of narrow spectra to inhomogeneous disturbances}

Alber's equation, which is a finite depth version of \eqref{eq: deep-water Alber}, has proved to be one of the main tools used to study the stability of ocean wave spectra to inhomogeneous disturbances. This is an important question, that has direct bearing on the suitability of modern wave-forecasting codes. To reiterate some main points: an ocean-wave spectrum represents a homogeneous, stationary sea-state, whose energy is transported at the group velocity according to linear theory. Nonlinear wave-wave interaction gives rise to a redistribution of energy from the middle frequencies to lower and higher frequencies, as well as changes in the frequencies (and thus velocities) of the waves themselves \cite{StuhlmeierStiassnie2019}. It should be borne in mind that this \emph{homogeneous} energy transfer acts on a timescale of order $T/\varepsilon^4,$ which for a typical period $T$ of 10 s, and a typical steepness of $\varepsilon=0.1$ works out to somewhat more than 27 hours. 

It is generally appreciated that statistical homogeneity is an idealization -- necessary for writing a time-independent spectrum theoretically, and required when measuring waves to establish such spectra at sea (see Hasselmann et al \cite[Sec.\ 2]{Hasselmann1973}\footnote{``Over 2000 wave spectra were measured; about [...] 121 corresponded to ``ideal" stationary and homogeneous wind conditions." p.\ 10.}). In light of this, it is important to establish whether even a small departure from homogeneity might invalidate the conclusions reached based on the homogeneous theory. The question addressed by Alber and others is exactly this: will inhomogeneities give rise to a faster energy exchange, and alter the energy distribution, and thus wave-statistics, of an otherwise homogeneous sea-state.

The case of unidirectional spectra has been particularly well studied, beginning with Alber \cite{Societyb}, and recent numerical and analytical work has shed light on many of the central issues. Approaches akin to Alber's, following the linear stability analysis \cite[Sec.\ 4]{Societyb} and arriving at his eigenvalue equation (4.16), have relied on integration (analytic in the case of simple spectral shapes like square, Gaussian, or Lorentz spectra in Stiassnie et al \cite{Stiassnie2008}, and numerical for more complex JONSWAP spectra) and parameter studies to establish instability criteria. Simply put, inhomogeneous disturbances will grow with the nonlinearity of the wave field, and with decreasing spectral bandwidth;  Gramstad \cite{Gramstad2017} has verified that unidirectional JONSWAP spectra are unstable for $\alpha \gamma/\varepsilon > 0.77,$ for $\varepsilon$ the mean wave slope, and $\alpha, \, \gamma$ the usual JONSWAP parameters, using Alber's criteria as well as Monte-Carlo simulations (using the Higher Order Spectral Method). A mathematically rigorous examination of the stability and instability of Alber's equation, including a study of well-posedness, was recently undertaken by Athanassoulis \emph{et al} \cite{Athanassoulis2018} for unidirectional spectra, putting earlier numerical results on solid footing.

For directional sea-states, the matter of instability was investigated by Ribal \emph{et al} \cite{Ribal2013}, who was able to extend earlier results for JONSWAP spectra to show that instability also depends on the degree of directional spreading -- narrower spectra again being more unstable.

\begin{figure}[H]
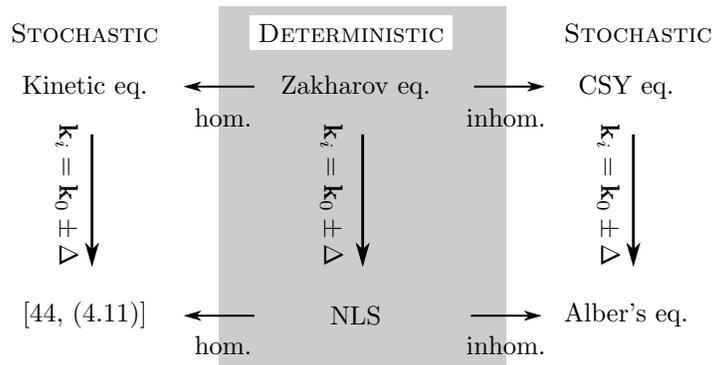

\centering
\begin{lpic}[l(20mm),r(20mm)]{Deterministic_Stochastic_Diagram(0.8)}
\lbl[c,W]{45,57;$\textsc{Deterministic}$}
\lbl[c]{0,57;$\textsc{Stochastic}$}
\lbl[c]{92,57;$\textsc{Stochastic}$}
\lbl[c]{45,48;$\text{Zakharov eq.}$}
\lbl[c]{0,48;$\text{Kinetic eq.}$}
\lbl[c]{90,48;$\text{CSY eq.}$}
\lbl[c]{45,10;$\text{NLS}$}
\lbl[c]{23,43;$\text{hom.}$}
\lbl[c]{23,5;$\text{hom.}$}
\lbl[c]{0,10;$\text{\cite[(4.11)]{Societyc}}$}
\lbl[c]{90,10;$\text{Alber's eq.}$}
\lbl[c]{70,43;$\text{inhom.}$}
\lbl[c]{70,5;$\text{inhom.}$}
\lbl[c]{-3,30,270;$\bk_i = \bk_0 \pm \Delta$}
\lbl[c]{41,30,270;$\bk_i = \bk_0 \pm \Delta$}
\lbl[c]{87,30,270;$\bk_i = \bk_0 \pm \Delta$}
\end{lpic}
\caption{Diagram of relationships between stochastic and deterministic equations. Indicated are assumptions of statistical homogeneity (``hom.") or inhomogeneity (``inhom."), as well as restrictions to narrow bandwidth ($\bk_i = \bk_0 \pm \Delta, \, |\Delta| \ll |\bk_0|$.)}
\end{figure}

From the deterministic perspective, the Benjamin-Feir instability derived from the NLS is an important mechanism in wave evolution. However, employing the Zakharov equation in place of the NLS \cite[Sec.\ VI.B, Fig.\ 23ff]{Yuen1982} yields a more realistic (finite) instability region. The same argument applies to the stochastic counterparts: instability should be investigated not only via the narrow-band NLS, but more generally for the Zakharov equation of which it is a special case. 

\subsubsection{Broad-band equations}
In the above sections, we have discussed stochastic evolution equations derived from the NLS, which implies a narrow bandwidth of order $\varepsilon k_0,$ for $k_0$ the carrier wave\-number. As mentioned at the beginning of the section, the equation which models nonlinear interaction in current wave-forecasting codes -- Hasselmann's kinetic equation (KE) -- has no such restriction. It is possible to derive this equation directly from the Zakharov equation \eqref{eq:3-Zakharov equation}, using the steps outlined in Section \ref{subsec:stochastic evolution equations}, but retaining terms up to sixth order in the moment hierarchy (see \cite[Eq.\ (2.6)ff]{Gramstad2013}). Further details can be found in \cite[Sec.\ 14.10]{Mei2005}, resulting in the equation 
\begin{align} \nonumber
\label{eq:KE}
\frac{dC(\bk,t)}{dt}=&4 \pi \iiint_{-\infty}^{\infty} T^2_{0,1,2,3} \left( C_2C_3(C_0 + C_1) - C_0 C_1(C_2 + C_3) \right)\\
& \delta(\bk + \bk_1 - \bk_2 - \bk_3) \delta(\omega + \omega_1 - \omega_2 - \omega_3) d\bk_1 d\bk_2 d\bk_3.
\end{align}
Here, as elsewhere, subscripts are understood to denote dependence on the wave\-number, so that e.g.\ $C_i = C(\bk_i,t).$ The kernel of the Zakharov equation appears again, and due to $\delta$ distributions in both wavenumber and frequency it follows that only exactly resonant quartets play a role in the interaction. It is also worth noting that \eqref{eq:KE} predicts no evolution for purely unidirectional waves -- symmetric quartets such as $\bk_a + \bk_b - \bk_a - \bk_b$ cause the integrand to vanish, and for nonsymmetric unidirectional quartets, the kernel $T$ vanishes \cite[p.\ 147]{Dyachenko1994}. This contrasts markedly with the narrow-banded case, where stochastic analogues of the (unidirectional) Benjamin-Feir instability play an important role.

\subsubsection{Stability of broad spectra to inhomogeneous disturbances}

A broad-banded evolution equation relaxing the assumption of spatial homogeneity was first derived by Crawford et al \cite{Crawford1980}, and recently studied for the case of a degenerate quartet of waves by Stuhlmeier \& Stiassnie \cite{Stuhlmeiere}. Like the kinetic equation, it is derived from the Zakharov equation \eqref{eq:3-Zakharov equation}, and due to the retention of the inhomogeneous terms it has a non-trivial evolution at the same order (and thus, the same time-scale) as the Alber equation (see \eqref{eq: deep-water Alber}). The discrete version of this equation, which is suitable for numerical computation, is
\begin{equation} \label{eq:4 CSY}
\frac{d r_{nm}}{dt} = i r_{nm} (\omega_m - \omega_n) + 2i \left( \sum_{p,q,r =1}^N T_{mpqr} r_{pq}r_{nr} \delta_{mp}^{qr} - \sum_{p,q,r =1}^N T_{npqr} r_{qp}r_{rm}\delta_{np}^{qr} \right).
\end{equation}

As mentioned above, underlying the idea of an energy spectrum (for a description of the ocean surface) is the property of statistical homogeneity. The energy spectrum thus consists of the homogeneous terms $r_{ii}$ only, and the equation \eqref{eq:4 CSY} provides a possibility to study whether such a spectrum undergoes some evolution if suitably perturbed. It is easy to note that if no inhomogeneous terms are present, i.e.\ the $r_{ij}$ vanish for $i \neq j,$ there is no evolution to this order -- the next order yields the KE \eqref{eq:KE}.

Let us write $r_{nm}$ as $r_{nm} = r^h_{nm}\delta_{nm} + \varepsilon r^i_{nm},$ where the superscripts $h$ and $i$ denote homogeneous and inhomogeneous terms, respectively. Substituting this into \eqref{eq:4 CSY} yields
\begin{equation}
\frac{d r^h_{nn}}{dt} = 0,
\end{equation}
and, for the inhomogeneous terms at order $\varepsilon$: 
\begin{align} \nonumber
\frac{1}{2i} \frac{d r_{nm}}{dt} = r_{nm} \left( \frac{\omega_m-\omega_n}{2} + \sum_p T_{mppm}r_{pp} - \sum_p T_{nppn} r_{pp} \right) \\ \label{eq:4 Instability Main Eq}
+ \sum_{p,q} T_{mpqn} r_{pq}r_{nn} \delta_{qn}^{mp} - \sum_{p,q} T_{npqm}r_{qp} r_{mm}\delta_{qm}^{np}.
\end{align}
This reduces for a degenerate quartet to the system studied by Stuhlmeier \& Stiassnie \cite{Stuhlmeiere}. Equation \eqref{eq:4 Instability Main Eq} describes a system of $n^2 - n$ linear, autonomous differential equations. For a given homogeneous initial state, the terms $r_{nn}$ are specified, and the system has the general form
\begin{equation}
\frac{1}{2i} \frac{d \mathbf{r}^i}{dt} = \mathbf{A} \mathbf{r}^i,
\end{equation}
for $\mathbf{A}$ the matrix of coefficients given in \eqref{eq:4 Instability Main Eq}, and $\mathbf{r}^i$ is the vector of the inhomogeneous correlators $r_{nm}, \, n \neq m.$ Negative eigenvalues of $\mathbf{A}$ therefore yield instability. That is, for a given homogeneous state, consisting of a specification of $N$ wave-vectors $\bk_1, \ldots, \bk_N$ and corresponding energy (or, equivalently, amplitude) in the form $r_{11}, \ldots, r_{NN},$ initially small inhomogeneous disturbances will grow exponentially with time and give rise to energy exchange within the framework of \eqref{eq:4 CSY}.

\begin{figure}[h]
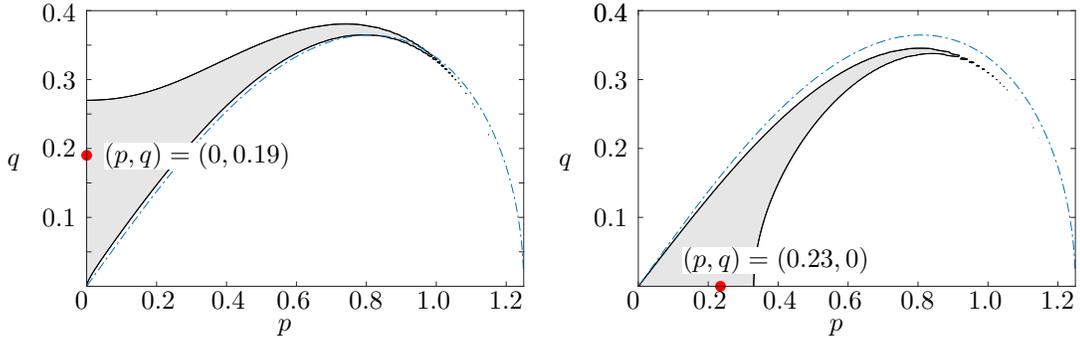

\centering
\begin{subfigure}{0.4\linewidth}
\begin{lpic}[]{Discriminant_Figure_1_nolabel(0.6)}
\lbl[t]{60,1;$p$}
\lbl[t]{0,37.5;$q$}
\lbl[t]{15,7;$0$}
\lbl[b]{10,21.5;$0.1$}
\lbl[b]{10,37;$0.2$}
\lbl[l,w]{20,37;$(p,q)=(0,0.19)$}
\lbl[b]{10,52;$0.3$}
\lbl[b]{10,67;$0.4$}
\lbl[t]{32,7;$0.2$}
\lbl[t]{47.5,7;$0.4$}
\lbl[t]{62,7;$0.6$}
\lbl[t]{78,7;$0.8$}
\lbl[t]{93,7;$1.0$}
\lbl[t]{109,7;$1.2$}
\end{lpic}
\end{subfigure}
\hspace{3em}
\begin{subfigure}{0.4\linewidth}
\begin{lpic}[]{Discriminant_Figure_2_nolabel(0.6)}
\lbl[t]{60,1;$p$}
\lbl[t]{0,37.5;$q$}
\lbl[t]{15,7;$0$}
\lbl[b]{10,21.5;$0.1$}
\lbl[b]{10,37;$0.2$}
\lbl[b]{10,52;$0.3$}
\lbl[b]{10,67;$0.4$}
\lbl[b]{10,67;$0.4$}
\lbl[t]{32,7;$0.2$}
\lbl[l,w]{26,14;$(p,q)=(0.23,0)$}
\lbl[t]{47.5,7;$0.4$}
\lbl[t]{62,7;$0.6$}
\lbl[t]{78,7;$0.8$}
\lbl[t]{93,7;$1.0$}
\lbl[t]{109,7;$1.2$}
\end{lpic}\end{subfigure}
\caption{Computed region of instability for \eqref{eq:4 CSY} (shaded region), for three waves $\bk_a = (1,0), \, \bk_b = (1+p,q), \, \bk_c = (1-p,-q),$ and for different wave slopes. Left panel: $\varepsilon_a = 0.01, \, \varepsilon_b = 0.1, \, \varepsilon_c = 0.1,$ the degenerate quartet with greatest growth rate (red dot) has $(p,q) = (0,0.19).$ Right panel: $\varepsilon_a = 0.1, \, \varepsilon_b = 0.01, \, \varepsilon_c = 0.01,$ the degenerate quartet with greatest growth rate (red dot) is collinear and has $(p,q) = (0.23,0).$}
\label{fig:Degenerate Quartet Discriminant}
\end{figure}

The case of a degenerate quartet $\bk_a = (1,0), \, \bk_b = (1+p,q), \, \bk_c = (1-p,-q),$ which satisfies $2\bk_a = \bk_b + \bk_c$  already demonstrates a range of possible behaviours. Two scenarios are presented in Figure \ref{fig:Degenerate Quartet Discriminant}, which depicts the domain of instability (shaded region) for different wave slopes.

For a sea-state with three random waves such that $(p,q)$ is in the shaded region of the figure, the evolution is changed by the presence of small inhomogeneities. One example of this subsequent evolution is given in Figure \ref{fig:CSY Evolution} below. The shaded grey region represents a ``warm-up" where the inhomogeneous terms (bottom panel) are small, and there is no evolution of the homogeneous terms (top panel). As this is an unstable case, the initially small inhomogeneities grow, and give rise to an energy exchange among the homogeneous terms. Further details on the choice of initial conditions, and the form of the inhomogeneities, may be found in \cite{Stuhlmeiere}.

\begin{figure}[ht]
\centering
\includegraphics[scale=0.7]{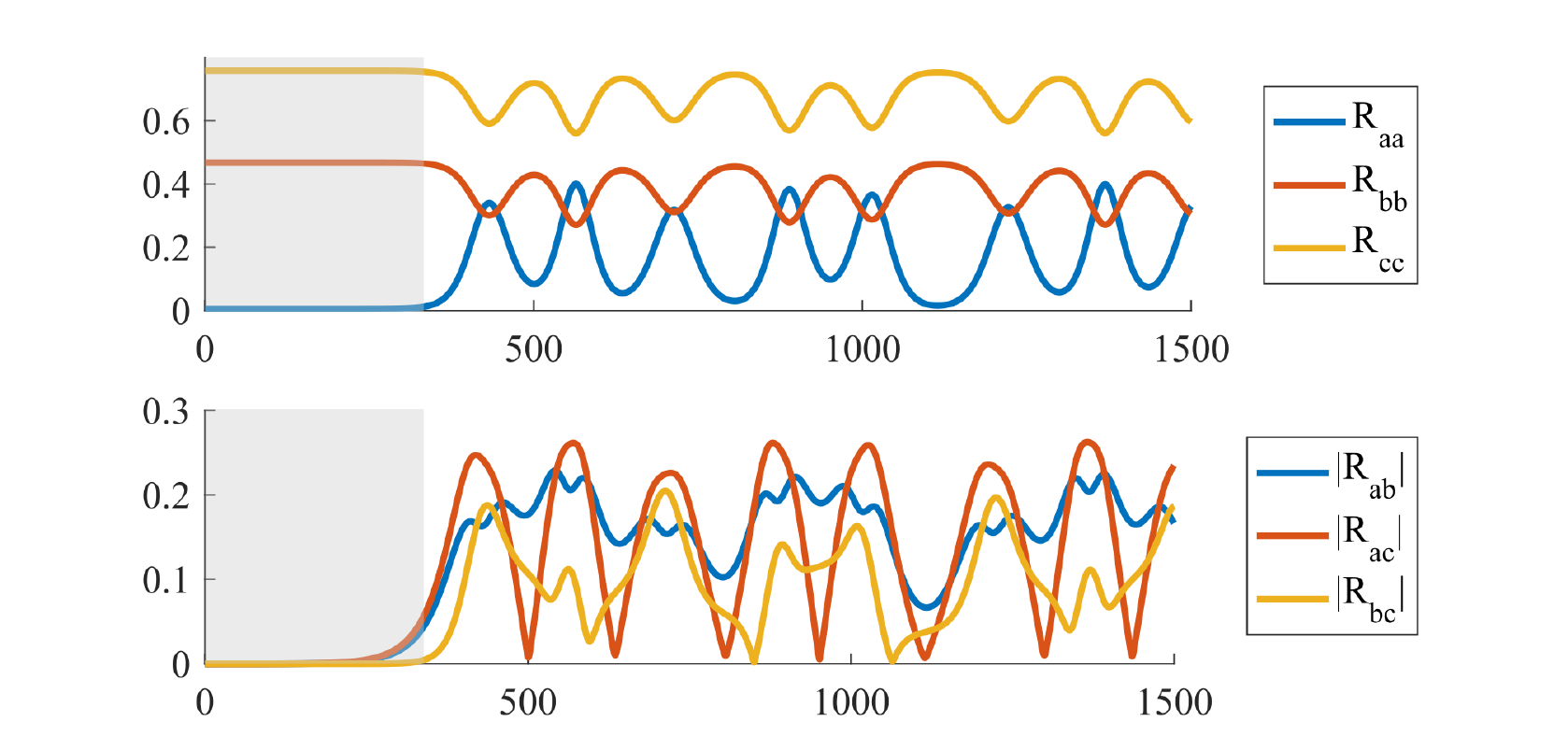}
\caption{Depiction of the evolution of a degenerate quartet $\bk_a = (1,0), \, \bk_b = (1.1,0.2), \bk_c=(0.9,-0.2)$ with $\varepsilon_a = 0.01, \, \varepsilon_b = 0.1, \, \varepsilon_c = 0.1,$ (see Figure \ref{fig:Degenerate Quartet Discriminant}, left panel) from time $t=0$ to $t=1500$ s, under small inhomogeneous disturbances. The top panel depicts the homogeneous terms, while the bottom panel depicts the magnitude of the inhomogeneous terms.}
\label{fig:CSY Evolution}
\end{figure}

More realistic cases, involving many modes, for which \eqref{eq:4 CSY} is a generalization of Alber's equation without a narrow-band restriction, await further study. 

\subsection{Stochastic evolution equations for coastal environments}

While deterministic (often called phase resolving) models can provide relatively
accurate calculations of the wave field evolution in coastal waters, they require
vast computational resources. This is due to the Nyquist limitation,
which enforces small grid spacing in a simulation, and due to a need
for a large number of runs required to obtain statistical quantities of interest. Indeed, running nonlinear deterministic models such as the nonlinear MSE \eqref{eq:NLMSE} or high-order Boussinesq (see, e.g., Madsen et al \cite{MadsenEtAl2006}) for large domains is a very computationally intensive procedure, which commonly reduces the range of practically calculated sea conditions and the size of the modelled region. Extending stochastic models to the nearshore can overcome this restriction by limiting such intensive, deterministic calculations to the very shallow region and the vicinity of coherent marine structures. In addition, it may allow for better nearshore wave forecasting capabilities. Therefore, the extension of stochastic models to the nearshore region is currently of great interest.

\subsubsection{One- and two-equation stochastic models in the nearshore region}

In the theory of waves in deep-water, cubic nonlinearities give rise naturally to equations for the spectrum  in terms of fourth order and sixth order averages (and respective cumulants). Just so, in the nearshore quadratic nonlinearities mean that the evolution of the spectrum is influenced by third order averages -- called the bispectra --  as investigated by Hasselmann et al \cite{hasselmann1963time}, Elgar \& Guza \cite{Elgar&Guza} and others. Two-equation nearshore stochastic models consist of an equation for the wave energy evolution (second order moment) with bi-spectral coupling terms, and another equation for the evolution of the bi-spectral components (see \cite{AgnonSheremet97,EldeberkyMadsen,K-HRasmussen1998,AgnonSheremet2000,VT2016}). Both equations are derived from the above deterministic models, and have the following general form:
\begin{align}
&\frac{\partial E_{p,l}}{\partial t_{1}}+\frac{\partial }{\partial x}\left
(C_{g,p} E_{p,l}\right)+2D_{p,l}C_{g,p}E_{p,l}\nonumber \\
& = -2 C_{g,p}
\sum_{\mathclap{\substack{{u=\max\{l-M,-M\}}\\{s=p-N}
}}}^{\mathclap{\substack{{s=N}\\
{u=\min\{l+M,M\}}}}}
\Re\left[\left(iW_{s,p-s,u,l-u}B_{s,u,p-s,l-u}\right)e^{-i\int\left(k_{s,u}^{x}+k_{p-s,l-u}^{x}-k_{p,l}^{x}\right)dx}\right],\label{eq:Stochastic}
\end{align}
which was derived from the deterministic equation \eqref{eq:EvolutionElevation} using the same procedure as in Sec.\ \ref{ssec:Narrow-band equations}: multiplying \eqref{eq:EvolutionElevation} by the complex conjugate of $a_{p,l}$, summing the result with its own complex conjugate, and applying an ensemble average. Here $\Re$ denotes the real part of an expression. The energy spectrum and bispectrum are defined as
\begin{eqnarray}
E_{p,l}=\left\langle \left|a_{p,l}\right|^{2}\right\rangle , &  & B_{s,u,p-s,l-u}=\left\langle a_{p,l}^{*}a_{s,u}a_{p-s,l-u}\right\rangle .\label{eq: E B}
\end{eqnarray}
The brackets $\left\langle \cdot \right\rangle $ denote the ensemble averaging
operation, the index $p$ defines the frequency of the spectral component, $l$ defines its lateral wavenumber, and the terms $D_{p,l},$ $W_{s,p-s,u,l-u}$ are defined as in Sec.\ \ref{ssec:Shallow water} (see \eqref{eq:EvolutionElevation}).

The bispectrum evolution equation is also derived from the same deterministic
model in a similar manner to yield
\begin{align}
& \frac{dB_{s,u,p-s,l-u}}{dx}+\left(D_{p,l}+D_{s,u}+D_{p-s,l-u}\right)B_{s,u,p-s,l-u}\nonumber \\
& +\left(\frac{C_{g,p}'}{2C_{g,p}}+\frac{C_{g,s}'}{2C_{g,s}}+\frac{C_{g,p-s}'}{2C_{g,p-s}}\right)B_{s,u,p-s,l-u}=\nonumber -i\left(I_{q,r,s-q,u-r,-p,l,p-s,l-u} \right. \nonumber \\
& T_{q,r,s-q,u-r,-p,l,p-s,l-u} +I_{q,r,s-q,u-r,-p,l,s,u}T_{q,r,s-q,u-r,-p,l,s,u} \nonumber \\  &\left.+I_{q,r,s-q,u-r,s,u,p-s,l-u}T_{q,r,s-q,u-r,s,u,p-s,l-u}\right) \label{eq:Bispectrum}
\end{align}
with the trispectrum components and their coefficients defined as
\begin{align} \label{eq:Trispectrum}
&T_{q,r,s-q,u-r,-p,l,p-s,l-u}=\left\langle a_{q,r}a_{s-q,u-r}a_{p,l}^{*}a_{p-s,l-u}\right\rangle,\\
& I_{q,r,s-q,u-r,-p,l,p-s,l-u}=
\sum_{\mathclap{\substack{{r=\max\{u-M,-M\}}\\{q=s-N}}}}^{\mathclap{\substack{{q=N}\\{r=\min\{u+M,M\}}}}}
W_{q,s-q,r,u-r}e^{-i\int\left(k_{q,r}^{x}+k_{s-q,u-r}^{x}-k_{s,u}^{x}\right)dx}.
\end{align}
Here, slow time changes of the spectral components were discarded, leading to a formulation of the bispectra as a function of only spatial coordinate.

In a similar manner it can be shown that the trispectrum will depend
on fourth order moments, which will in turn depend on fifth order
moments and so forth. Therefore, for solving the system a closure
relation is required, as when deriving \eqref{eq: deep-water Alber}. A quasi-Gaussian closure \cite{Benney&Saffman1966} is
applied to truncate the infinite hierarchy of equations, resulting in 
\begin{align}
& \frac{dB_{s,u,p-s,l-u}}{dx}+\left(D_{p,l}+D_{s,u}+D_{p-s,l-u}\right)B_{s,u,p-s,l-u}\nonumber \\
& +\left(\frac{C{}_{g,p}'}{2C{}_{g,p}}+\frac{C{}_{g,s}'}{2C{}_{g,s}}+\frac{C{}_{g,p-s}'}{2C{}_{g,p-s}}\right)B_{s,u,p-s,l-u} \nonumber \\
& =-2i\left(W{}_{-s,-(p-s),u,l-u}E_{s,u}E_{p-s,l-u}\right.\nonumber \\
& \left.+W{}_{p,-s,l,l-u}E_{p,l}E_{s,u}+W{}_{p,-(p-s),l,l-u}E_{p,l}E_{p-s,l-u}\right)e^{i\int\left(k_{s,u}^{x}+k_{p-s,l-u}^{x}-k_{p,l}^{x}\right)dx}.\label{eq:EvolutionBenney}
\end{align}
Equations (\ref{eq:Stochastic}) and (\ref{eq:EvolutionBenney}) comprise
a two-equation stochastic model, which can be used to solve the shoaling
problem.

The number of permutations between wave components constructing the
various bi-spectral components is very large, so that the problem becomes very computationally intensive. In order to address this
limitation, equation (\ref{eq:EvolutionBenney}) is solved for $B_{s,u,p-s,l-u}$.
The bispectrum is assumed to be negligible in deep water as in this
region the sea is nearly Gaussian. Applying the integrating factor
method to equation (\ref{eq:EvolutionBenney}) yields an analytical solution for
the bispectrum, which can then be substituted into \eqref{eq:Stochastic} to construct a one-equation model (see \cite{AgnonSheremet97,AgnonSheremet2000,VT2016}) of the form
\begin{align}
&\frac{\partial E_{p,l}}{\partial t_{1}}+\frac{\partial }{\partial x}\left
(C_{g,p} E_{p,l}\right)+2D_{p,l}C_{g,p}E_{p,l} \\
& =  4 C_{g,p} \sum_{\mathclap{\substack{{u=\max\{l-M,-M\}}\\{s=p-N}}}}^{\mathclap{\substack{{s=N}\\{u=\min\{l+M,M\}}}}}
 \Re\left[Q_{s,p-s,u,l-u}+Q_{p,s,l,t-u}+Q_{p,p-s,l,l-u}\right]W_{s,p-s,u,l-u},\nonumber \\
\label{eq:one equation model}
\end{align}
with
\begin{align}
 Q_{s,p-s,u,l-u} & =  e^{-i\int_{0}^{x}\left(k_{s,u}^{x'}+k_{p-s,l-u}^{x'}-k_{p,l}^{x'}\right)dx'}\label{eq:Q definition} {e^{-\int_{0}^{x}-J_{s,u,p-s,l-u}dx'}}\\
&  \int_{0}^{x} \left( E_{s,u}E_{p-s,l-u}W{}_{-s,-(p-s),u,l-u}e^{i\int_{0}^{x'}\left(k_{s,u}^{x''}+k_{p-s,l-u}^{x''}-k_{p,l}^{x''}\right)dx''} \right. \nonumber \\
& \left. \times e^{\int_{0}^{x'}-J_{s,u,p-s,l-u}dx''} \right) dx'.\nonumber 
\end{align}
where the $J$-term represents summation of all linear coefficients.

In order to simplify the calculation of the $Q$-terms, as a first approximation, one can assume slow spectral evolution with respect to shoaling coefficients and take the energy terms outside of the integral (see \cite{AgnonSheremet97,AgnonSheremet2000}), similar to the procedure adopted when deriving the kinetic equation \eqref{eq:KE} for the evolution of spectra in deep water (see \cite[Eq.\ (A1)ff]{Gramstad2013}). This may reduce the accuracy in breaking regions where wave heights change significantly within short distances.

\subsection{Localization procedures for nearshore stochastic models}

The one-equation model \eqref{eq:one equation model} reduces the number of equations to be solved significantly. However, the solution still requires the calculation of non-local nonlinear coefficients. This makes its implementation difficult for operational models based on the wave-action equation (WAE). Furthermore, the evaluation of the bispectrum  \eqref{eq:Bispectrum} or the non-local coefficients \eqref{eq:Q definition} still enforces a strict Nyquist limitation as they themselves may oscillate quite rapidly in space.  A localisation of the $Q$ coefficient can therefore allow for operational model implementation while improving  the efficiency of calculation significantly.

Simplified approaches are currently employed in source terms used in operational wave models (see \cite{EldeberkyBattjes,BecqEtAl1999,salmon2016consistent}). They approximate the bi-spectral evolution equation using empirical data for the second harmonic, or assume very small changes in the bi-spectra to solve it as an algebraic equation.

A more advanced localization approach, which aims to further improve operational models in the nearshore region, was first considered in Stiassnie \& Drimer \cite{Stiassnie&Drimer} and subsequently improved in Toledo \& Agnon \cite{ToledoAgnonK} and Vrecica \& Toledo \cite{VT2016}.
This procedure entails extracting spectral components out of the integral, and only
the mean part of the bispectra -- the main interest in such models -- is accounted for. Assuming a monotone
slope, the $Q$-term is simplified from \eqref{eq:Q definition} as
\begin{equation} \label{eq: Q localized}
Q_{s,p-s,u,l-u}\left(x\right)=E_{s,u}E_{p-s,l-u}P_{s,p-s,u,l-u}\left(h\left(x\right),h'\left(x\right)\right),
\end{equation}
which enables pre-calculation of the nonlinear coupling term $P$, which is a localized coefficient.

\begin{figure}
	\centering
\includegraphics[width = 0.9\textwidth]{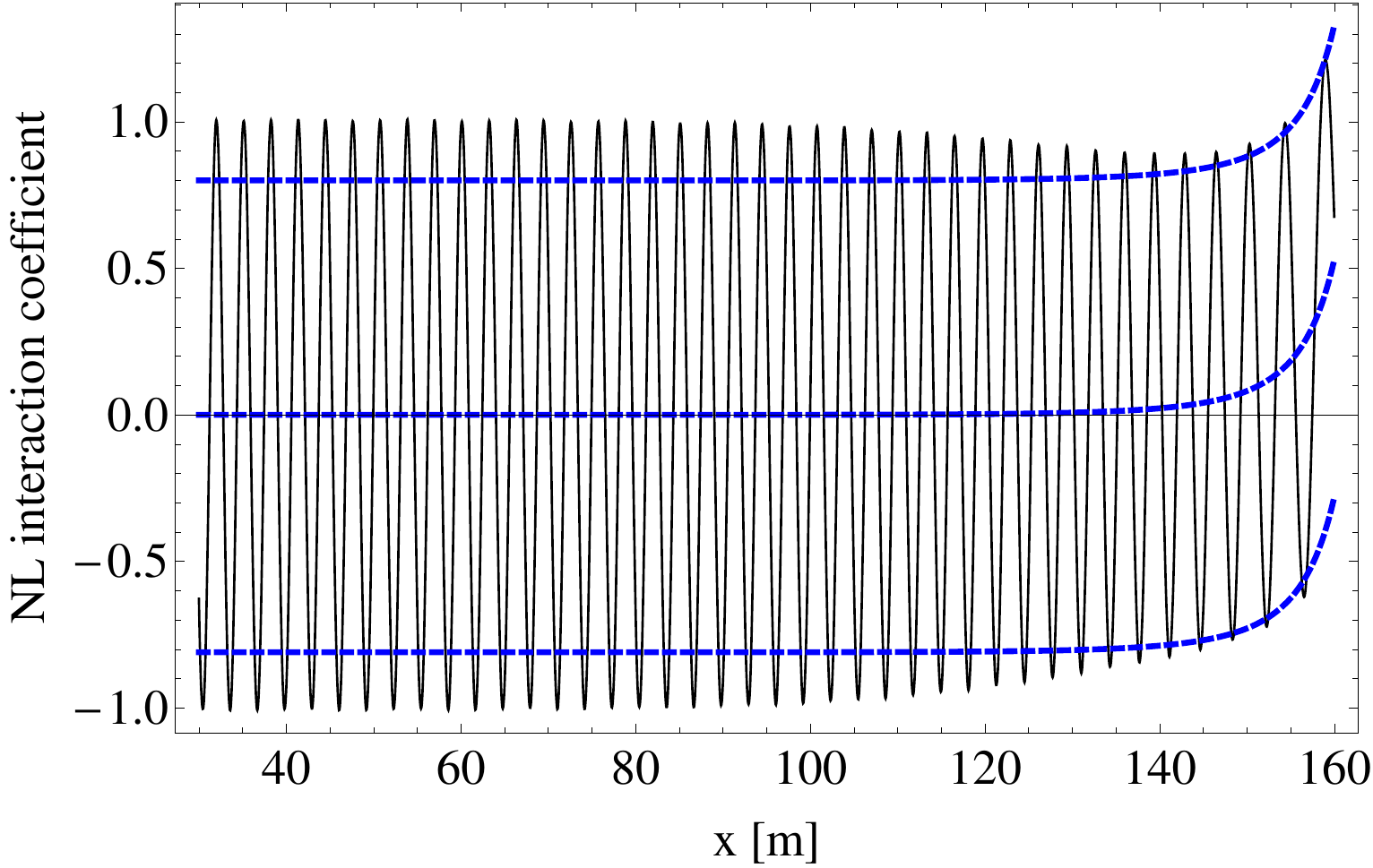}
	\caption{\label{fig:Localizations}Two different behaviors of the bispectra for the case of monochromatic wave propagation over a slope, as shown in \cite{VT2016}.
		In intermediate waters the nonlinear coupling term ($Q_{1100}$) is mostly oscillating with a slowly growing mean component (left panel). As the water gets shallower ($kh$ reduces)
		the Q-term tends to an asymptotic shallow water solution (right panel).}
	
\end{figure}

The different behaviours of the non-local nonlinear interaction coefficients \eqref{eq:Q definition} and the localized simplification \eqref{eq: Q localized} can be seen
in figure \ref{fig:Localizations} for a monochromatic wave energy transfer to the second harmonic while shoaling on a beach with a constant slope. In deep water, the nonlinear interaction term (or the bispectra) oscillates in space with no mean change (see figure \ref{fig:Localizations} (left panel) in the region of $x<80$m). This indicates a bound wave behaviour with mean energy transfer between the modes. Once the wave enters intermediate depths, class III triad resonance conditions \eqref{eq:triad resonance} can be satisfied,
and the nonlinear interaction coefficient oscillates, albeit with a small mean change. This indicates a mean energy transfer to the second harmonic in the class III Bragg resonance mechanism as shown in figure \ref{fig:Resonance}a. 

As all the interacting waves enter shallow water conditions, they asymptotically go to resonance as in figure \ref{fig:Resonance}c, so instead of an oscillatory behaviour they act in an exponential manner (see figure \ref{fig:Localizations}, right panel). Under such conditions, a distinct localization approach is needed compared to the intermediate water case. Such an approach was developed in Vrecica \& Toledo \cite[Sec.\ 4.3.2]{VT2016}. The two different formulations of the localized model are separated by a gate term, which is a function of depth and bottom slope (see figure \ref{fig:Localizations}). It is stressed that
the transition between these two behaviours also depends on the bottom slope, for sharp changes the shallow water localization will activate sooner, and vice versa.

A limitation of the model is its inapplicability to long propagation in shallow water with strong nonlinearity. Energy would cascade to ever higher frequencies, leading to wave breaking. This condition is more relaxed for cases with large directional spreading, as waves are still dispersive in the angular space (see \cite{Newell&Aucoin}). However, operational stochastic models are usually not extended to such areas, typically a Bousinessq-type model (e.g, \cite{MadsenEtAl2006}), or RANS model (\cite{SWASH}) would be used for such cases.

\subsection{Comparison between deterministic ensembles and stochastic equations\label{subsec:Statistics}}

In deep water waves are often considered to be uncorrelated, and the slow evolution of the spectral energy density can be captured by e.g.\ the kinetic equation \eqref{eq:KE}. However, as waves propagate
to nearshore correlations build up, and coherent patterns form \cite{Newell2011}. Therefore, a quasi-Gaussian closure, while commonly used in nearshore wave models, is often not valid for cases involving strong nonlinearity or dissipation. Such quasi-Gaussian closure can result in an overestimate of energy transfer to higher frequencies, as well as result in (unphysical) negative energies (see \cite[Sec.\ 4.4]{TimJanssenPhD}). The closure of Holloway \cite{Holloway1980}, which adds dissipation to the bispectral evolution equation, is often used as an empirical solution.

Nonlinear shoaling also affects the wave shape, which is commonly expressed using skewness
and asymmetry, as discussed in Elgar \& Guza \cite{Elgar&Guza}. Initially, in deep water, wave skewness (which relates
to wave phase), and asymmetry (which relates to nonlinear energy transfer)
are both near zero. As the wave field starts shoaling both begin to grow,
however in the surf zone skewness tends back to zero, while the limit
for asymmetry is $\sim3$. These depend on the value of the Ursell number $a \lambda^2/h^3$ (for $a, \, \lambda, \, h$ a typical sea-surface elevation, horizontal length scale, and vertical length scale, respectively), as shown in \cite{FreilichEtAl1990}. 

\begin{figure}
\includegraphics[width = 0.45 \textwidth]{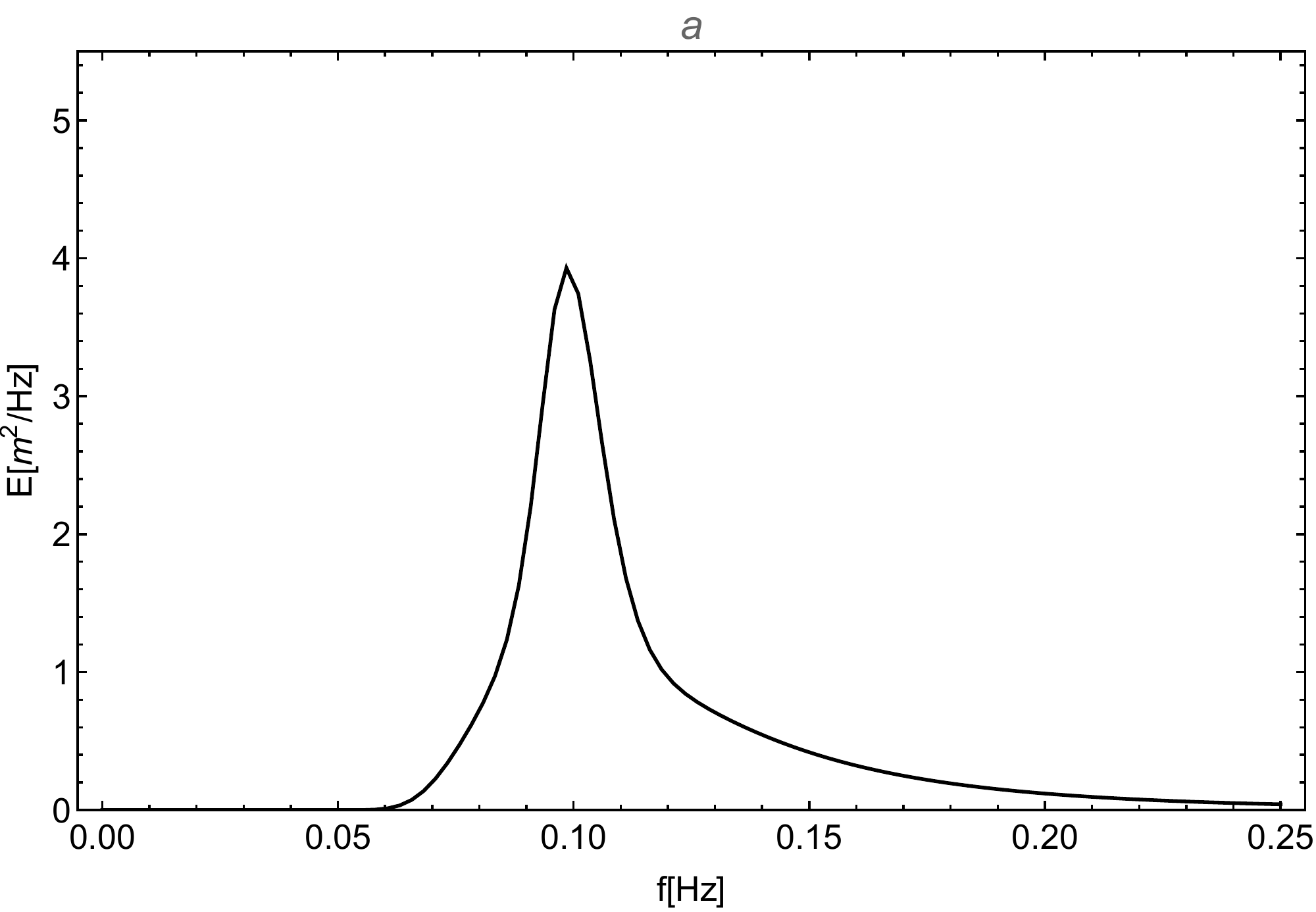}$\quad$\includegraphics[width = 0.45 \textwidth]{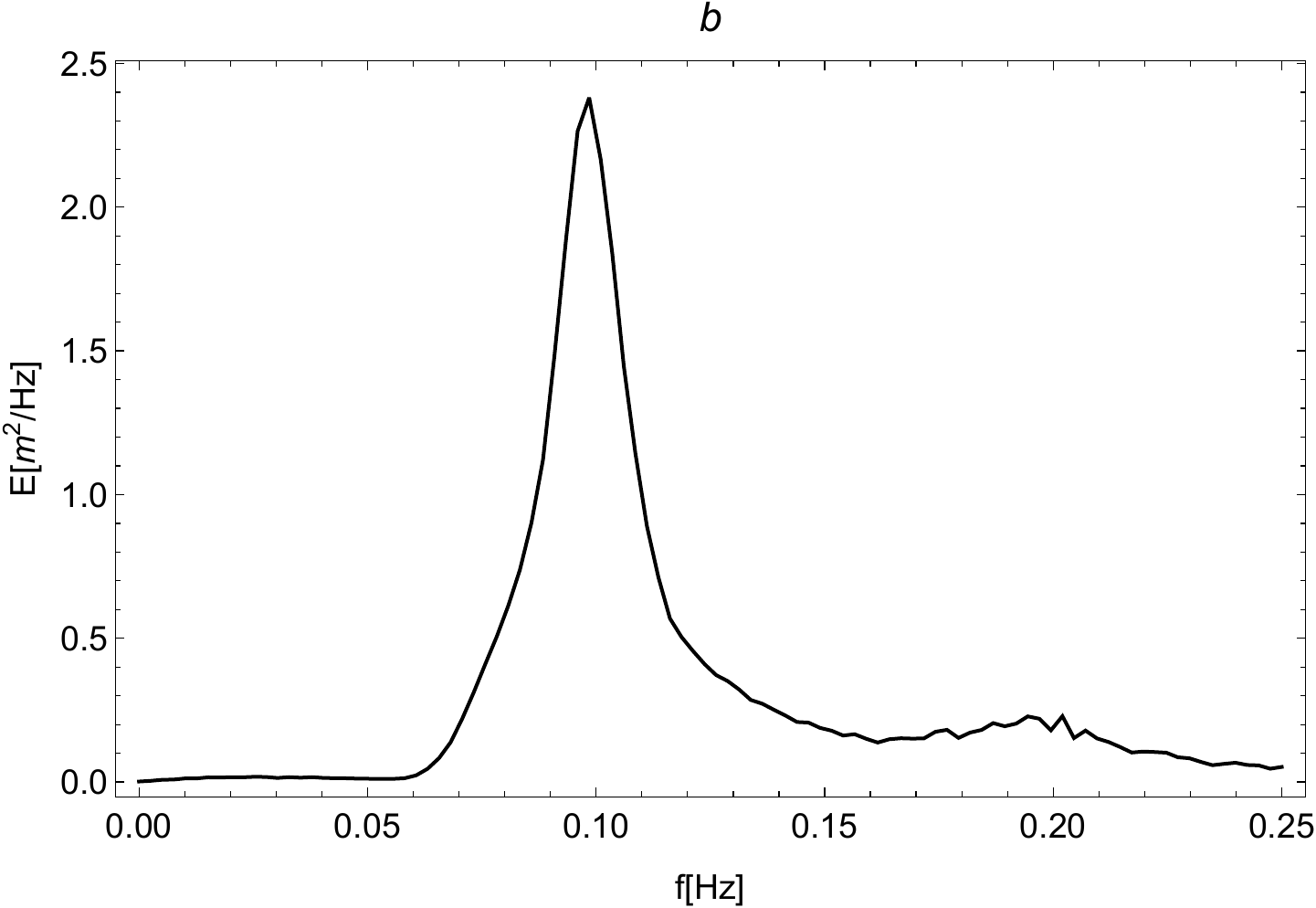}

\includegraphics[width = 0.45 \textwidth]{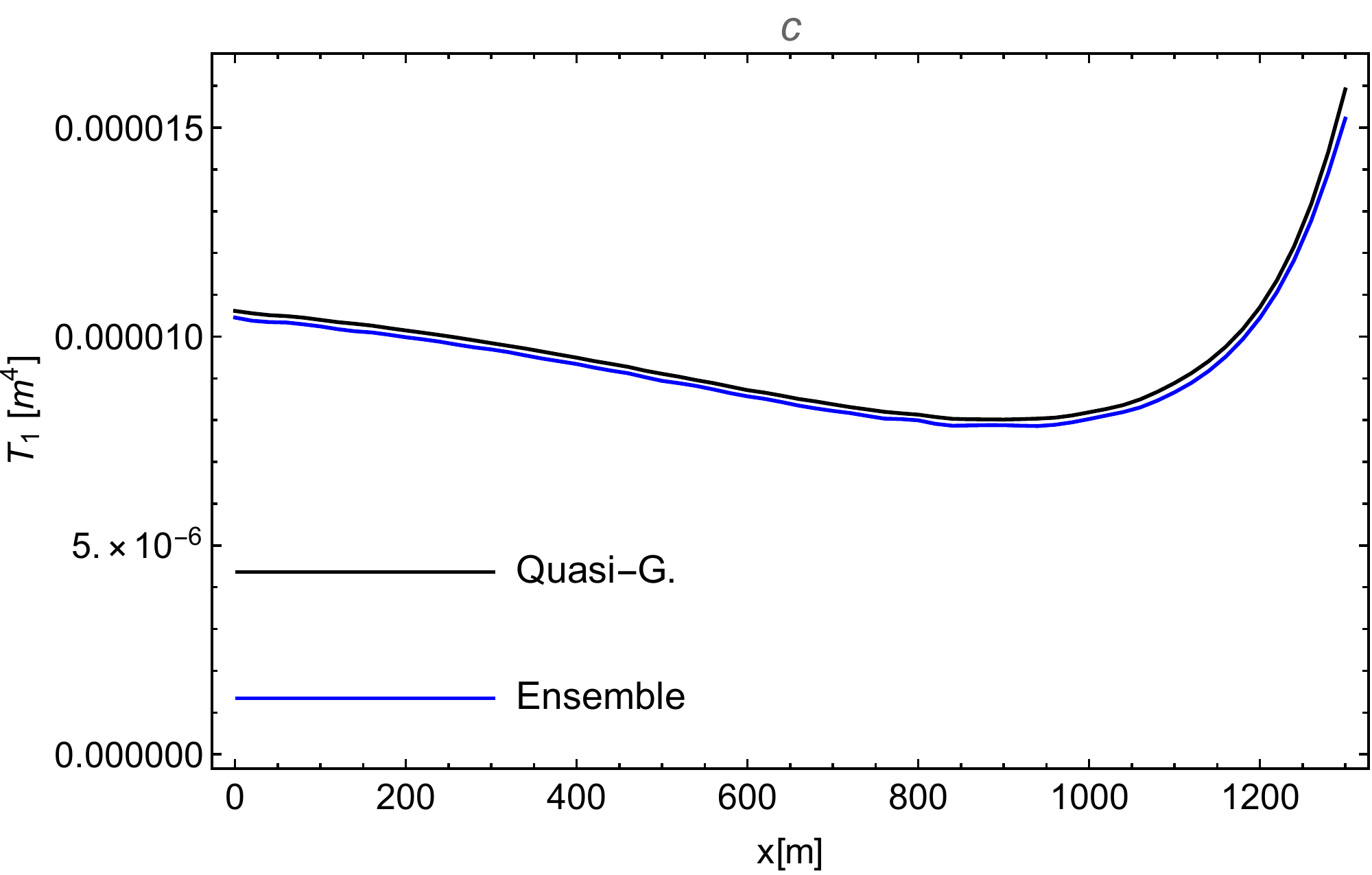}$\quad$\includegraphics[width = 0.45 \textwidth]{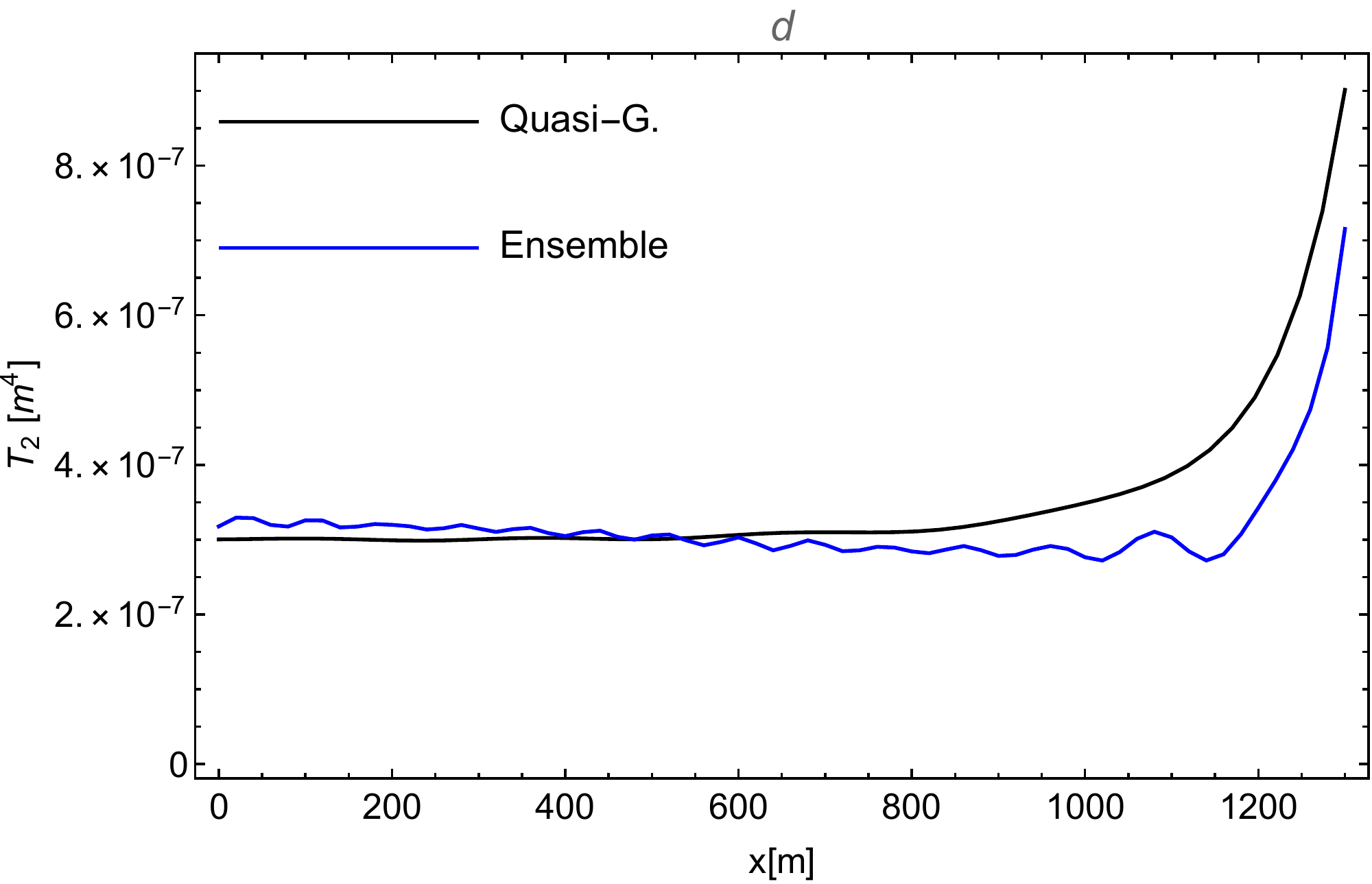}

	\caption{\label{fig:Spectrum}Evolution of a JONSWAP spectra, obtained using equation \eqref{eq:EvolutionElevation}, over a monotone
		(5\%) slope from deep water (a) to 5m depth (b). Comparison of quasi-Gaussian, and ensemble
		averaged trispectral moments ($T_{1}=T_{39,41,-q,-80+q}$ and $T_{2}=T_{-39,80,-q,-41+q}$, lateral indices are dropped for brevity) at 5m depth, describing energy transfer to secondary peak (c), and backtransfer to the primary peak (d).}
\end{figure}

 In order to shed light on the validity of the quasi-Gaussian closure a simulation of unidirectional JONSWAP spectra with a 1 m significant wave height and peak frequency of 0.1 Hz was performed using the nonlinear MSE
\eqref{eq:EvolutionElevation}. The spectrum propagates from deep water (\ref{fig:Spectrum}a) to 5 m depth (\ref{fig:Spectrum}b) over a five percent slope.  In the present case, the Ursell number is small, so skewness and asymmetry may be neglected. The quasi-Gaussian closure for trispectral moments describing energy transfer to the secondary peak ($E_{39,0}$ and $E_{41,0}$ to $E_{80,0}$, as defined by equation \ref{eq: E B} is compared against ensemble averaged trispectral sums. Comparison is also made between trispectral moments describing backtransfer of energy to $E_{41,0}$. The quasi-Gaussian moments are defined using equation \ref{eq:Trispectrum} as $T_{1}=E_{39,0}E_{41,0}$ and $T_{2}=E_{39,0}E_{80,0}$, while the ensemble averaged ones are defined as $T_{1}=\sum_{q=-N}^{N}a_{39,0}a_{41,0}a_{q}^{*}a_{80-q,0}^{*}$, and $T_{2}=\sum_{q=-N}^{N}a_{39,0}^{*}a_{80,0}a_{q}^{*}a_{41-q,0}^{*}$  for the super- and sub-harmonic interactions respectively ($N$ represents the number of discretized wave harmonics). The results are shown in figures \ref{fig:Spectrum}c and \ref{fig:Spectrum}d.

As a relatively mild nonlinear case is considered, the quasi-Gaussian closure \cite{Benney&Saffman1966} is accurate to leading order. Based on preliminary analysis, the closure is accurate for the trispectral moments containing the most energetic wave components up to this point. However, the errors are not proportional to each trispectrum, and the relative error is larger for less energetic components. When summed over all possible indices the errors can become significant in cases of strong nonlinearity.

\begin{figure}
\includegraphics[width = 0.45 \textwidth]{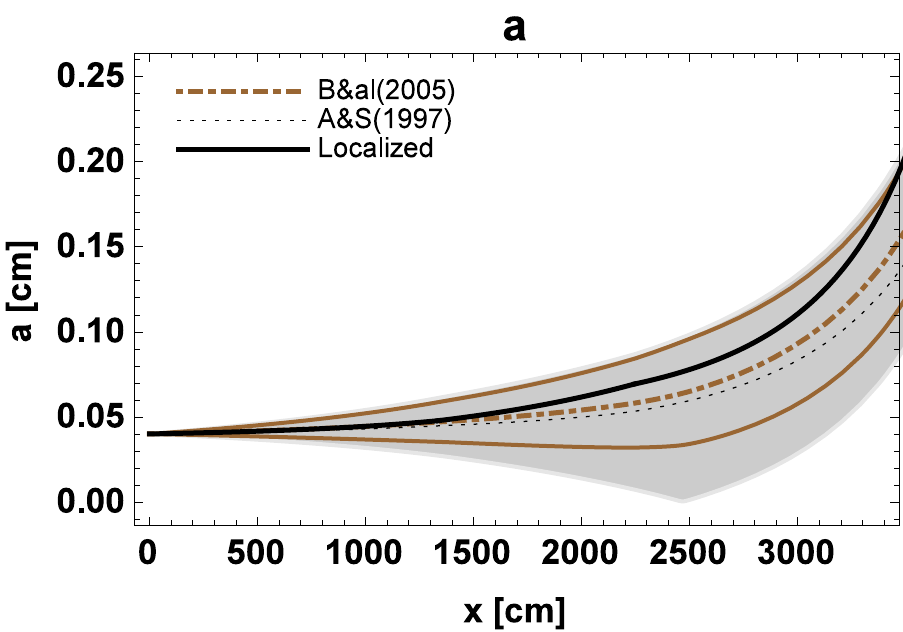}$\quad$\includegraphics[width = 0.45 \textwidth]{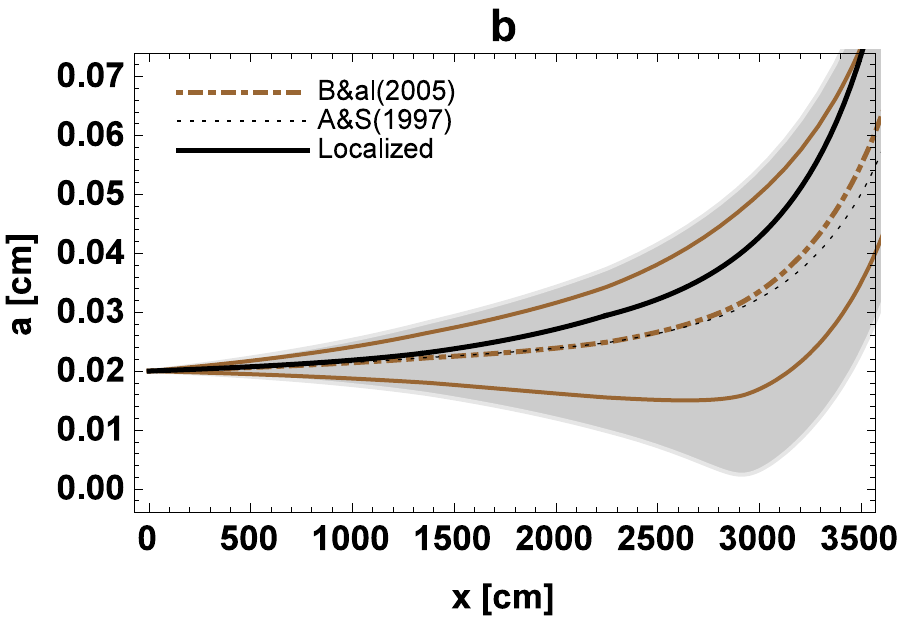}
	\caption{\label{fig:Statistics} Evolution of a 
		low-frequency wave (0.01Hz) of initial amplitudes 0.04 cm (a) and 0.02 cm
		(b) due sub-harmonic interaction of bi-chromatic waves (0.11 Hz and 0.1 Hz) over a monotone beach. Grey area and dot-dashed brown line: Monte-Carlo realizations using equation \eqref{eq:EvolutionElevation} and their ensemble average. Brown
		lines: the ensemble's standard deviation added and subtracted from the averaged
		result. Thick black line: localized stochastic model.  Dotted black line: the ensemble averaged deterministic result of
		\cite{AgnonSheremet97}.}
	
\end{figure}

While averaged equations based on the quasi-Gaussian closure may agree well with ensemble averages of the deterministic equations, it is important to point out that individual realizations of the deterministic equations may show significant deviations from the average. This is illustrated via the generation of infragravity waves (0.01Hz), where equation
\eqref{eq:EvolutionElevation} is solved with an  input of a bichromatic wave field (0.1 and 0.11 Hz, with amplitudes of 2.07 cm for figure \ref{fig:Statistics}a and 1.22 cm for figure \ref{fig:Statistics}b). The ensemble relates to different relative phases between the bi-chromatic waves. An envelope containing all realizations is shown in grey  together with their mean value and standard deviation. It can be clearly seen that depending on the relative phase of interacting components, each realization can be drastically different. Hence, it should be taken into account that the ensemble averaged wave height may be significantly lower than that of the largest realization. 

\section{Conclusions \& Perspectives}
We have focused throughout on the mechanism of resonant (or near-resonant) energy exchange, and how it drives wave evolution in deep water as well as finite depth. 
A variety of deterministic and stochastic model equations exist, suitable for simulating the evolution of a wave field from deep to shallow water. Deterministic models are able to give insight through many realizations, while stochastic models are much faster, enabling analysis of larger areas. 
Although present day operational wave models are highly reliable overall, there is considerable work to be done on the theoretical front. In particular, while an accuracy within a few percent for a wave forecast may be suitable for the vast majority of situations, it is often the outliers, or extreme events (see, e.g.\ Adcock \& Taylor \cite{Adcock2014} for a review), that have the most potential to cause damage. We highlight a few areas where further study is needed for deep and shallow water.

\subsection{Direct numerical simulations and kinetic equations}
While it is a formally simple step to average an evolution equation, the relationship between solutions of the deterministic equation and its random counterpart is a subtle one, as highlighted above in Section \ref{subsec:Statistics}. Ideally, we would observe the following: if averaging is done by assuming, say, phases randomly and independently distributed over $(0,2\pi],$ then the behaviour of the averaged equation should be the average of many realizations of the deterministic equation, where each realization chooses random phases independently distributed over $(0,2\pi].$ This is Monte-Carlo simulation, where computationally a ``realization'' means a solution with given initial data. 

The simplest nontrivial case already shows some difficulties: a resonant quartet of four waves. The results of Stiassnie \& Shemer \cite[Sec.\ 9, Fig.\ 8]{Stiassnie2005} and Annenkov \& Shrira \cite[Sec.\ 3, Fig.\ 2]{Annenkov2006} demonstrate clear discrepancies. In both cases four initial conditions $b(\bk_1,t=0),\ldots b(\bk_4,t=0)$ are supplied. In the Monte-Carlo simulation, the Zakharov equation is integrated when $b(\bk_1,0)=|b(\bk_1,0)|e^{i\phi}$ for many realizations with different $\phi\in(0,2\pi].$ The kinetic equation is integrated with the (phase-free) initial condition $C_i(0)=|b(\bk_i,0)|^2.$ Annenkov \& Shrira \cite[Fig.\ 3]{Annenkov2006} were able to obtain good agreement with the kinetic equation only after replacing each of the four waves in the Monte-Carlo simulation with a cluster of five waves, with a resulting 181 coupled quartets.

The comparison between numerical simulation and the kinetic equation for JONSWAP spectra with many modes has also been extensively explored. Despite the fact that the kinetic equation is derived with some long-time asymptotic limits (to eliminate non-resonant contributions), and is formally on the long time-scale $T_4,$ Tanaka \cite{Tanaka2007} found that it captures spectral changes for a broad ($\cos^2(\theta)$) JONSWAP spectrum on a much shorter time-scale. Tanaka's results point to the fact that ensemble averaging is inessential provided the mode density is high enough -- which is related to the theoretical view of the kinetic theory as averaging out redundancy in the dynamic description \cite[Sec.\ 2.1.2]{Zakharov1992a} -- but the four-wave results point to a lower limit to the applicability of the kinetic theory.

Direct numerical simulations can also be used to investigate other averaged equations. For example, the instability results of Alber \cite{Societyb} were compared with Monte-Carlo simulations of the NLS by Onorato \emph{et al} \cite[Sec.\ 3]{Onorato2003} for unidirectional Lorentz spectra, and some qualitative agreement with the theoretical instability region was found. For initial Gaussian spectra, Dysthe \emph{et al} \cite{DYSTHE2003} found approximate agreement between Alber's results and Monte-Carlo simulations only in the unidirectional case -- for directional Gaussian spectra there were marked discrepancies. All these results, as well as the recent study by Annenkov \& Shrira \cite{Annenkov2018} comparing several wave kinetic equations with Monte-Carlo simulations, point to a need for intense further study.

\subsection{The role of near-resonant interactions}
The classical kinetic equation \eqref{eq:KE} contains $\delta$-functions in both wavenumber and frequency -- this reflects its derivation under the assumptions of exact resonance. This is the consequence of an asymptotic limit, described in detail by Janssen \cite[see Eq.\ 27]{Janssen2003}. The Zakharov equation requires no such assumption, and includes near-resonant interactions such that $\Delta_{n+p-q-r} = \omega_n + \omega_p - \omega_q - \omega_r$ is of order $\varepsilon^2.$ Indeed, the side-band instability relies on exactly these (fast, compared to the kinetic time-scale) interactions.

A number of generalizations of the kinetic equation \eqref{eq:KE} exist, which aim to incorporate near-resonant interactions. Janssen \cite[Eq.\ 25]{Janssen2003} proposed one such equation, Annenkov \& Shrira \cite[Eq.\ 2.25]{Annenkov2006} proposed another, and Gramstad \& Stiassnie \cite[Eq.\ 2.19]{Gramstad2013} generalized this to include frequency correction terms. The same generalized equation was recently studied by Andrade et al \cite{Andrade2019}, and found to exhibit finite-time blow-up for some degenerate quartets.  When performing Monte-Carlo simulations, Annenkov \& Shrira \cite[Sec. 3.4]{Annenkov2006} found that omitting waves in exact resonance, and keeping only their near-resonant neighbors had no effect on the subsequent evolution. There are clearly many more nearly-resonant interactions than there are exactly resonant ones. This fact, together with the need to discretize in wavenumber-space when performing computations, means that some amount of coarse-graining is inevitable. Such near resonant generalizations should be explored in detail, and compared with the kinetic equation (see Annenkov \& Shrira \cite{Annenkov2018} for the initiation of such an effort).

\subsection{Nearshore wave modelling}

Much work remains to be done on important aspects of coastal wave modelling. 
The stochastic nonlinear formulation used in the breaking region (see Section \ref{subsec:Statistics}) has limitations, as the quasi-Gaussian closure it employs is not valid in the surf zone. The empirical closure of \cite{Holloway1980} corrects overestimation of energy transfer, but does not fully describe nearshore wave statistics. Hence, an in-depth study of nearshore wave statistics is still required in order to formulate better stochastic closures. Of particular note is the fact that an arbitrary realization of the deterministic solution can drastically vary from the ensemble averaged result. This can be seen even in the very simplistic case of a subharmonic interaction of bi-chromatic waves (see figure \ref{fig:Statistics}).  For the evolution of a JONSWAP spectrum (figure \ref{fig:Spectrum}), which has numerous phase selection possibilities, is was necessary to generate 200 realizations in order to obtain good agreement between the quasi-Gaussian closure and ensemble averaged trispectra.

Another nearshore phenomenon to consider is wave reflection. Incident waves will be scattered backwards from the bottom slope (see \cite{Yevnin&Toledo} for the linear case) and nonlinearly generated in the backwards direction. In addition, most models assume linear dispersion for every wave harmonic (\cite{JanssenEtAl2006} is good counterexample), and neglect formation of coherent wave patterns, but how important these are is yet to be investigated.

Wave breaking is one of the most important mechanisms in the nearshore region. Its formulations are empirical by nature (usually based on \cite{BattjesJanssen1978}). Therefore, they better resolve cases close to the ones under which they were tested, and commonly require tuning of their coefficients (see \cite{MaseKirby1992, EldeberkyBattjes1996}). As most laboratory measurements are conducted in wave flumes, the resulting formulations are limited to directly incident waves without addressing two dimensional aspects.  Furthermore, when nearshore nonlinear interactions are not well described in the models used for their formulations, these breaking formulations may not separate well between the two mechanisms. Hence, they have problems representing the complex combined behaviour of breaking and nonlinear interaction, and may require changes and recalibration for any advancement in the nonlinear modelling.

\subsection*{Acknowledgment}
 RS is grateful for the hospitality and support of the Erwin Schr\"odinger Institute for Mathematics and Physics (ESI), Vienna, Austria, as well as support from a Small Grant from the IMA.

\begin{multicols}{2}

\end{multicols}

\end{document}